\documentclass[aps,pra,preprint]{revtex4-1}
\usepackage{amsfonts,bm,graphicx,hyperref,physics,tikz,upgreek}
\usepackage[utf8]{inputenc}
\usepackage{BOONDOX-cal}
\usepackage[caption=false]{subfig}
\usetikzlibrary{arrows.meta,shapes}
\usetikzlibrary{calc,positioning}
\newcommand{\ce}{\ensuremath{\textrm{e}}}
\newcommand{\ci}{\ensuremath{\textrm{i}}}
\newcommand{\cpi}{\ensuremath{\uppi}}

\usepackage{color}

\begin{document}
\title{Realizing Quantum Boltzmann Machines Through Eigenstate Thermalization}
\author{Eric R.\ Anschuetz}
\email{eans@mit.edu}
\affiliation{MIT Center for Theoretical Physics, 77 Massachusetts Avenue, Cambridge, MA 02139, USA}
\affiliation{Zapata Computing, Inc., 501 Massachusetts Avenue, Cambridge, MA 02139, USA}
\author{Yudong Cao}
\email{yudong@zapatacomputing.com}
\affiliation{Zapata Computing, Inc., 501 Massachusetts Avenue, Cambridge, MA 02139, USA}
\preprint{MIT-CTP/5106}

\begin{abstract}
    Quantum Boltzmann machines are natural quantum generalizations of Boltzmann machines that are expected to be more expressive than their classical counterparts, as evidenced both numerically for small systems and asymptotically under various complexity theoretic assumptions. However, training quantum Boltzmann machines using gradient-based methods requires sampling observables in quantum thermal distributions, a problem that is \textsf{NP}-hard. In this work, we find that the locality of the gradient observables gives rise to an efficient sampling method based on the Eigenstate Thermalization Hypothesis, and thus through Hamiltonian simulation an efficient method for training quantum Boltzmann machines on near-term quantum devices. Furthermore, under realistic assumptions on the moments of the data distribution to be modeled, the distribution sampled using our algorithm is approximately the same as that of an ideal quantum Boltzmann machine. We demonstrate numerically that under the proposed training scheme, quantum Boltzmann machines capture multimodal Bernoulli distributions better than classical restricted Boltzmann machines with the same connectivity structure. We also provide numerical results on the robustness of our training scheme with respect to noise.
\end{abstract}

\maketitle

\section{Introduction}

Boltzmann machines are one of the earliest neural network architectures in classical machine learning~\cite{ACKLEY1985147}, and have been used in both supervised and unsupervised learning settings. They serve as a versatile tool for learning real-world data distributions. In essence, a classical Boltzmann machine is a thermal set of spins that interact under an Ising Hamiltonian, which is diagonal in the natural basis of states representing the combination of spin-ups and spin-downs of the system.

Recent efforts in quantum computation have unveiled that by considering quantum Hamiltonians which are non-diagonal and under certain natural complexity theoretic assumptions, one is able to perform learning tasks more efficiently than with classical computation~\cite{Amin2018,Kieferova2017,2019arXiv190205162W}. However, an outstanding practical question is then how to efficiently sample gradient observables in quantum thermal states, a necessary condition for the efficiency of training quantum Boltzmann machines. Some proposals in the literature point toward using quantum annealing devices as sources of these thermal states, though due to the challenges in controlling the interaction between the quantum annealer and its external thermal bath, the thermalization process often ``freezes out'' before thermal equilibrium is established~\cite{PhysRevA.92.052323,Amin2018}. Furthermore, efforts in using quantum annealers for Boltzmann machines are often challenged by the inherent noise, connectivity, and form of coupling allowed in the annealing device~\cite{Mandra2016StrengthsApproaches}. 

On gate-model quantum computers, variational methods for producing classical thermal states have been proposed~\cite{Verdon2017ACircuits,2018arXiv181111756W} using the Quantum Approximate Optimization Algorithm~\cite{Farhi2014}, though for an $N$-spin system these methods require $N$ ancilla qubits; furthermore,~\cite{Verdon2017ACircuits} only considers diagonal Hamiltonians. Other variational methods require no ancilla qubits~\cite{2018arXiv180403023M,2018arXiv181201015M}, though such approaches often require retraining the ansatz for different thermal states, which may be costly when many thermal states must be sampled from in succession as in the training of quantum Boltzmann machines. Nonvariational methods include generalizations of classical Monte Carlo methods to the quantum regime~\cite{PhysRevLett.103.220502,Yung754,2016arXiv160302940N}, but may not be practical on near-term quantum devices. Finally, there are classes of proposed methods that rely on various assumptions on the underlying quantum system, such as taking a mean field approximation~\cite{2015arXiv150702642W} or relying on short correlation lengths in the underlying system~\cite{2019arXiv190107653M}, neither of which are guaranteed to hold in classical or quantum Boltzmann machines.

Here, we present a heuristic method that allows one to prepare a pure state which locally approximates a (potentially highly correlated) quantum thermal state at a known temperature using only $\mathcal{O}\left(1\right)$ ancilla qubits and time evolution under a chaotic, tunable quantum Hamiltonian. Our construction is supported by the Eigenstate Thermalization Hypothesis (ETH)~\cite{PhysRevE.50.888,PhysRevA.43.2046,DAlessio2016}, which is a statement about how subsystems of pure states thermalize under certain conditions. Although analytical results on the subject are sparse, ETH has been substantiated in a broad set of regimes both numerically~\cite{rigol2008thermalization,PhysRevE.93.032104,Garrison2018} and experimentally~\cite{neill2016ergodic,Kaufman794}. By utilizing chaotic quantum Hamiltonians in our quantum Boltzmann machines, we are able to perform a quantum quench procedure to locally sample observables in quantum thermal states. Furthermore, our scheme is set up such that there is a method of approximately obtaining the inverse temperature of the system, which is needed for estimating the correct sign and magnitude of the gradient of the quantum Boltzmann machine loss function.

The remainder of this paper is organized as the following: Sec.~\ref{sec:bm} describes the basic setting of Boltzmann machines, both classical and quantum. Sec.~\ref{sec:eth} describes the Eigenstate Thermalization Hypothesis, its conditions, and what it predicts. Sec.~\ref{sec:thermal} describes our thermal state preparation protocol, and Sec.~\ref{sec:ns} demonstrates numerical simulations of our procedure. Finally, we conclude and discuss future work in Sec.~\ref{sec:conclusion}.

\section{Boltzmann Machines}\label{sec:bm}

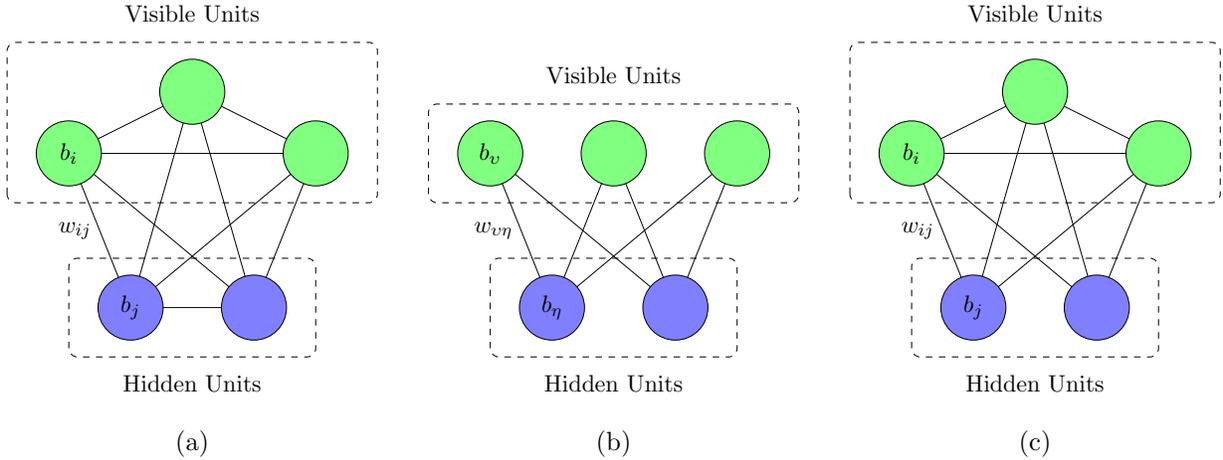
\begin{figure}[ht]
    \begin{center}
        \begin{subfloat}[\label{fig:bm}]{\resizebox{0.3\linewidth}{!}{\begin{tikzpicture}
            \draw (-1, -5) -- node[anchor=east]{$w_{ij}$} (-2, -2.5);
            \draw (-1, -5) -- (0, -1.5);
            \draw (-1, -5) -- (2, -2.5);
            \draw (1, -5) -- (-2, -2.5);
            \draw (1, -5) -- (0, -1.5);
            \draw (1, -5) -- (2, -2.5);
            \draw (-2, -2.5) -- (0, -1.5);
            \draw (-2, -2.5) -- (2, -2.5);
            \draw (0, -1.5) -- (2, -2.5);
            \draw (-1, -5) -- (1, -5);
            \node[draw,circle,fill=blue!50,minimum size=30,inner sep=0] at (-1, -5) {$b_j$};
            \node[draw,circle,fill=blue!50,minimum size=30,inner sep=0] at (1, -5) {};
            \draw[dashed,rounded corners] (-2, -5.8) rectangle (2, -4.2) {};
            \node[anchor=south] at (0, -6.5) {Hidden Units};
            \node[draw,circle,fill=green!50,minimum size=30,inner sep=0] at (-2, -2.5) {$b_i$};
            \node[draw,circle,fill=green!50,minimum size=30,inner sep=0] at (0, -1.5) {};
            \node[draw,circle,fill=green!50,minimum size=30,inner sep=0] at (2, -2.5) {};
            \draw[dashed,rounded corners] (-3, -3.3) rectangle (3, -0.7) {};
            \node[anchor=south] at (0, -0.5) {Visible Units};
            \end{tikzpicture}}}
        \end{subfloat}
        \hspace{0.025\linewidth}
        \begin{subfloat}[\label{fig:rbm}]{\resizebox{0.3\linewidth}{!}{\begin{tikzpicture}
            \draw (-1, -5) -- node[anchor=east]{$w_{\upsilon\eta}$} (-2, -2.5);
            \draw (-1, -5) -- (0, -2.5);
            \draw (-1, -5) -- (2, -2.5);
            \draw (1, -5) -- (-2, -2.5);
            \draw (1, -5) -- (0, -2.5);
            \draw (1, -5) -- (2, -2.5);
            \node[draw,circle,fill=blue!50,minimum size=30,inner sep=0] at (-1, -5) {$b_{\eta}$};
            \node[draw,circle,fill=blue!50,minimum size=30,inner sep=0] at (1, -5) {};
            \draw[dashed,rounded corners] (-2, -5.8) rectangle (2, -4.2) {};
            \node[anchor=south] at (0, -6.5) {Hidden Units};
            \node[draw,circle,fill=green!50,minimum size=30,inner sep=0] at (-2, -2.5) {$b_{\upsilon}$};
            \node[draw,circle,fill=green!50,minimum size=30,inner sep=0] at (0, -2.5) {};
            \node[draw,circle,fill=green!50,minimum size=30,inner sep=0] at (2, -2.5) {};
            \draw[dashed,rounded corners] (-3, -3.3) rectangle (3, -1.7) {};
            \node[anchor=south] at (0, -1.5) {Visible Units};
            \end{tikzpicture}}}
        \end{subfloat}
        \hspace{0.025\linewidth}
        \begin{subfloat}[\label{fig:srqbm}]{\resizebox{0.3\linewidth}{!}{\begin{tikzpicture}
            \draw (-1, -5) -- node[anchor=east]{$w_{ij}$} (-2, -2.5);
            \draw (-1, -5) -- (0, -1.5);
            \draw (-1, -5) -- (2, -2.5);
            \draw (1, -5) -- (-2, -2.5);
            \draw (1, -5) -- (0, -1.5);
            \draw (1, -5) -- (2, -2.5);
            \draw (-2, -2.5) -- (0, -1.5);
            \draw (-2, -2.5) -- (2, -2.5);
            \draw (0, -1.5) -- (2, -2.5);
            \node[draw,circle,fill=blue!50,minimum size=30,inner sep=0] at (-1, -5) {$b_j$};
            \node[draw,circle,fill=blue!50,minimum size=30,inner sep=0] at (1, -5) {};
            \draw[dashed,rounded corners] (-2, -5.8) rectangle (2, -4.2) {};
            \node[anchor=south] at (0, -6.5) {Hidden Units};
            \node[draw,circle,fill=green!50,minimum size=30,inner sep=0] at (-2, -2.5) {$b_i$};
            \node[draw,circle,fill=green!50,minimum size=30,inner sep=0] at (0, -1.5) {};
            \node[draw,circle,fill=green!50,minimum size=30,inner sep=0] at (2, -2.5) {};
            \draw[dashed,rounded corners] (-3, -3.3) rectangle (3, -0.7) {};
            \node[anchor=south] at (0, -0.5) {Visible Units};
            \end{tikzpicture}}}
        \end{subfloat}
        \caption{(a) An example Boltzmann machine. The units are coupled with interaction weights $w_{ij}$. Each unit also has a local bias field $b_i$. (b) An example RBM. Samples are drawn from the visible units, and correlations between visible units are created through couplings with the hidden layer. The visible units are coupled with the hidden units through interaction weights $w_{\upsilon\eta}$. Each unit also has a local bias field $b_i$. (c) An example QBM with a semi-restricted architecture. The units are coupled with interaction weights $w_{ij}$. Each unit also has a local bias field $b_i$. Furthermore, off-diagonal fields and interactions are included in the Hamiltonian (see Sec.~\ref{sec:bm}).
        \label{fig:full_rbm}}
    \end{center}
\end{figure}

The goal of generative modeling in machine learning is to train a model that generates data points that resemble a given set of data. In particular, the \emph{Boltzmann machine} is an energy-based generative model that models the given data set as a thermal state under the classical Ising energy function
\begin{equation}
    E\left(\bm{z};\bm{\theta}\right)=\sum\limits_i b_i z_i+\sum\limits_{i,j} w_{ij} z_i z_j,
\end{equation}
where $\bm{z}\in\left\{-1,1\right\}^n$ is a binary vector and $\bm{\theta}=\left\{\bm{b},\bm{w}\right\}$ are the model parameters. In practice, the spins are separated into a bipartite structure of \emph{visible units} and \emph{hidden units} such that approximate sampling of the visible units of these thermal states can be performed through Gibbs sampling~\cite{doi:10.1162/089976602760128018}. To make this structure explicit by labeling the $n_v$ visible units with $\upsilon$ indices and the $n_h$ hidden units with $\eta$ indices, the energy function is of the form:

\begin{equation}
    E\left(\bm{z};\bm{\theta}\right)=\sum\limits_i b_i z_i+\sum\limits_{\upsilon,\eta} w_{\upsilon\eta} z_\upsilon z_\eta.
\end{equation}
Boltzmann machines with this internal structure are termed \emph{restricted Boltzmann machines} (RBMs)---a visual comparison between a general Boltzmann machine (Fig.~\ref{fig:bm}) and an RBM (Fig.~\ref{fig:rbm}) is given in Fig.~\ref{fig:full_rbm}.

\emph{Quantum Boltzmann machines} (QBMs) are natural quantum generalizations of classical Boltzmann machines~\cite{Amin2018}. They are described by a quantum thermal probability distribution
\begin{equation}
p_\beta\left(\bm{z_v};\bm{\theta}\right)=\frac{\tr\left(\varPi_{\bm{z}_v}\ce^{-\beta H_{\textrm{QBM}}\left(\bm{\theta}\right)}\right)}{\tr\left(\ce^{-\beta H_{\textrm{QBM}}\left(\bm{\theta}\right)}\right)}
\end{equation}
and, as introduced in~\cite{Amin2018}, defined by a semi-restricted transverse Ising Hamiltonian
\begin{equation}
H_{\textrm{QBM}}\left(\bm{\theta}\right)=\sum\limits_i\varGamma_i\sigma_i^x+\sum\limits_i b_i\sigma_i^z+\sum\limits_{\upsilon,i} w_{\upsilon i}\sigma_\upsilon^z\sigma_i^z,
\label{eq:transverse_ising}
\end{equation}
%
%
where $\bm{z_v}\in\left\{-1,1\right\}^{n_v}$, $\bm{\theta}=\left\{\bm{\varGamma},\bm{b},\bm{w}\right\}$, and $\varPi_{\bm{z_v}}$ is a projector of the visible units of the QBM onto $\bm{z_v}$. Here, semi-restricted means that the only disallowed connections are between the hidden units---a visual representation is given in Fig.~\ref{fig:srqbm}. In general, one could consider a more general Hamiltonian given by~\cite{Kieferova2017}:
\begin{equation}
H_{\textrm{QBM}}\left(\bm{\theta}\right)=H_{\textrm{off-diag}}\left(\bm{\theta}_{\textrm{off-diag}}\right)+\sum\limits_i b_i\sigma_i^z+\sum\limits_{\upsilon,i} w_{\upsilon i}\sigma_\upsilon^z\sigma_i^z,
\end{equation}
where $H_{\textrm{off-diag}}\left(\bm{\theta}_{\textrm{off-diag}}\right)$ is composed of terms that are not diagonal in the computational basis. For instance, taking $H_{\textrm{off-diag}}$ to be composed of tunable $\sigma_i^x$ and $\sigma_i^x\sigma_j^x$ terms makes the ground state problem of $H\left(\bm{\theta}\right)$ \textsf{QMA}-complete, and therefore is generally believed to be more expressive than the Hamiltonian of Eq.~\eqref{eq:transverse_ising}, which is generally believed to not be \textsf{QMA}-complete~\cite{PhysRevA.78.012352}. We also consider QBMs with the same connectivity as RBMs; see Appendix~\ref{sec:systems} for the details of all Hamiltonian models we consider.

In both the classical and quantum cases, the parameters $\bm{\theta}$ are trained such that the negative log-likelihood
\begin{equation}
\mathcal{L}\left(\bm{\theta}\right)=-\sum\limits_{\bm{z_v}}p_{\textrm{data}}\left(\bm{z_v}\right)\log\left(p_\beta\left(\bm{z_v};\bm{\theta}\right)\right)
\label{eq:exact_ll}
\end{equation}
is minimized, where $p_\beta\left(\bm{z_v};\bm{\theta}\right)$ is the thermal distribution corresponding to either a classical Boltzmann machine or a QBM. For QBMs, gradients of $\mathcal{L}$ are not efficiently samplable; thus, in practice, one trains on an upper bound of the loss function given by~\cite{Amin2018}:
\begin{equation}
\tilde{\mathcal{L}}\left(\bm{\theta}\right)=-\sum\limits_{\bm{z_v}}p_{\textrm{data}}\left(\bm{z_v}\right)\log\left(\frac{\tr\left(e^{-\beta H_{\bm{v}}\left(\bm{\theta}\right)}\right)}{\tr\left(e^{-\beta H_{\textrm{QBM}}\left(\bm{\theta}\right)}\right)}\right),
\label{eq:u_b_ll}
\end{equation}
where
\begin{equation}
H_{\bm{z_v}}\left(\bm{\theta}\right)=H_{\textrm{QBM}}\left(\bm{\theta}\right)-\ln\left(\varPi_{\bm{v}}\right).
\end{equation}
Training a QBM on $\tilde{\mathcal{L}}$ not only generally prevents finding the optimal QBM parameters for the true loss function $\mathcal{L}$, but also makes training $\bm{\theta}_{\textrm{off-diag}}$ generally impossible~\cite{Amin2018}. Using clever generalized measurements it is possible to train these off-diagonal elements, though deriving such measurements requires prior knowledge of the data distribution and thus is generally difficult in practice~\cite{Kieferova2017}. In this work, we only consider training on the upper bound $\tilde{\mathcal{L}}$ of the true loss function $\mathcal{L}$. We note that it is also possible to train QBMs on a relative entropy loss function~\cite{Kieferova2017,2019arXiv190205162W}, but we do not explore that method in our work.
%
%

For a generic QBM, derivatives of Eq.~\eqref{eq:u_b_ll} with respect to the diagonal parameters $\left\{\bm{b},\bm{w}\right\}$ are of the form:
%
%
\begin{align}
    \partial_{b_i}\tilde{\mathcal{L}}\left(\bm{\theta}\right)&=\beta\sum\limits_{\bm{z_v}}p_{\textrm{data}}\left(\bm{z_v}\right)\left(\frac{\tr\left(\sigma_i^z \ce^{-\beta H_{\bm{z_v}}\left(\bm{\theta}\right)}\right)}{\tr\left(\ce^{-\beta H_{\bm{z_v}}\left(\bm{\theta}\right)}\right)}-\frac{\tr\left(\sigma_i^z\ce^{-\beta H_{\textrm{QBM}}\left(\bm{\theta}\right)}\right)}{\tr\left(\ce^{-\beta H_{\textrm{QBM}}\left(\bm{\theta}\right)}\right)}\right),\label{eq:b_grad}\\
    \partial_{w_{ij}}\tilde{\mathcal{L}}\left(\bm{\theta}\right)&=\beta\sum\limits_{\bm{z_v}}p_{\textrm{data}}\left(\bm{z_v}\right)\left(\frac{\tr\left(\sigma_i^z \sigma_j^z\ce^{-\beta H_{\bm{z_v}}\left(\bm{\theta}\right)}\right)}{\tr\left(\ce^{-\beta H_{\bm{z_v}}\left(\bm{\theta}\right)}\right)}-\frac{\tr\left(\sigma_i^z \sigma_j^z \ce^{-\beta H_{\textrm{QBM}}\left(\bm{\theta}\right)}\right)}{\tr\left(\ce^{-\beta H_{\textrm{QBM}}\left(\bm{\theta}\right)}\right)}\right).\label{eq:w_grad}
\end{align}
For an observable $O_\theta$ corresponding to the $\theta$-component of the gradient, this can be equivalently expressed as:
\begin{equation}
    \partial_\theta\tilde{\mathcal{L}}\left(\bm{\theta}\right)=\beta\left(\mathbb{E}_{\bm{z_v}\sim p_{\textrm{data}}}\left[\left\langle O_\theta\right\rangle_{\bm{z_v}}\right]-\langle O_\theta\rangle_\text{QBM}\right),
    \label{eq:grad}
\end{equation}
where the first expectation value is averaged with respect to the data distribution and the second with respect to the model distribution. Due to the form of $H_{\bm{v}}$, the first term of these derivatives---the \emph{positive phase}---is efficiently computable classically~\cite{Amin2018}. The second term---the \emph{negative phase}---is not believed to be efficiently computable in general, and if done exactly would require sampling from a general quantum thermal distribution, which is \textsf{NP}-hard~\cite{1982JPhA...15.3241B}. Our main contribution is developing a practical heuristic method for approximately sampling the local observables of Eq.~\eqref{eq:grad} from this quantum thermal distribution, taking advantage of the low weight of the operators that must be sampled.

\section{Local Quantum Thermalization}\label{sec:eth}

\subsection{The Eigenstate Thermalization Hypothesis}\label{sec:real_eth}

%
%
A necessary prerequisite of training QBM states is being able to sample local observables from thermal states at a known temperature. In general, preparing such thermal states is \textsf{NP}-hard~\cite{1982JPhA...15.3241B}. However, isolated quantum systems are known to thermalize locally; the mechanism under which this is believed to occur is known as the \emph{Eigenstate Thermalization Hypothesis} (ETH)~\cite{PhysRevA.43.2046,PhysRevE.50.888,DAlessio2016}. ETH states that subsystem thermalization occurs on the level of eigenstates of the system; namely, it gives an ansatz for the matrix elements of observables in the eigenbasis $\left\{\ket{E_i}\right\}$ of the Hamiltonian~\cite{Srednicki_1999,DAlessio2016}:
\begin{equation}
    \bra{E_j}O\ket{E_i}=O_\omega\left(\overline{E}\right)\delta_{ij}+\ce^{-\frac{S\left(\overline{E}\right)}{2}}f_O\left(\overline{E},E_i-E_j\right)R_{ij}.
    \label{eq:eth_ansatz}
\end{equation}
Here, $E_i=\bra{E_i}H\ket{E_i}$ and $\overline{E}=\frac{E_i+E_j}{2}$ is the average energy. $O_\omega\left(\overline{E}\right)$ is the expectation value of the microncanonical ensemble at an energy $\overline{E}$ with an energy window $\omega$, which is given by:
\begin{equation}
    O_\omega\left(\overline{E}\right)=\sum\limits_{E'\in\left[\overline{E}-\frac{\omega}{2},\overline{E}+\frac{\omega}{2}\right]}\ket{E'}\bra{E'},
\end{equation}
where $\omega$ vanishes in the thermodynamic limit (that is, as the system size $n_v$ is taken to infinity; usually, $\omega$ is taken to be $\mathcal{O}\left(\frac{\overline{E}}{\sqrt{n_v}}\right)$).
%
%
Finally, $S$ is the microcanonical entropy, $f_O$ is a smooth function, and $R_{ij}$ is a complex random variable with zero mean and unit variance. Though unproven analytically, this ansatz is conjectured to hold for all operators with support on less than half of the system in nonintegrable systems~\cite{Garrison2018}.

Furthermore, in the thermodynamic limit if an operator $O$ has equal microcanonical and canonical expectation values given a Hamiltonian $H$, the following holds:
\begin{equation}
    \bra{E}O\ket{E}=\frac{\tr\left(O\ce^{-\beta\left(E\right)H}\right)}{\tr\left(\ce^{-\beta\left(E\right)H}\right)},
    \label{eq:canonical}
\end{equation}
where $\beta\left(E\right)$ is such that:
\begin{equation}
    E=\frac{\tr\left(H\ce^{-\beta\left(E\right)H}\right)}{\tr\left(\ce^{-\beta\left(E\right)H}\right)}.
    \label{eq:e_beta_rel}
\end{equation}
The microcanonical and canonical ensembles generically agree on the expectation values of observables with a volume of support sublinear in $n$ for nonintegrable systems in the thermodynamic limit (assuming the entropy is concave in $\overline{E}$, which is typical in most physical settings)~\cite{TOUCHETTE2004138,Ellis2000}.

In short, Eq.~\eqref{eq:canonical} is expected to hold for all observables with a volume of support sublinear in $n$ in the thermodynamic limit~\footnote{Though this is guaranteed only asymptotically, numerically we find this to be true even for relatively small system sizes (see Sec.~\ref{sec:ns}).}, for all systems that exhibit an equivalence between the microcanonical and canonical ensembles~\cite{PhysRevE.97.012140,Garrison2018}. For systems that do not thermalize in the conventional sense, such as integrable systems and many-body localized systems, this equivalence is generalized to an equivalence between the microcanonical ensemble and the generalized canonical ensemble~\cite{Costeniuc2005,PhysRevLett.111.127201,DAlessio2016}; we leave the consideration of such systems to future work.

\subsection{Quantum Quench Dynamics and Sampling Local Observables}\label{sec:quench}

Given that ETH holds in the sense of Eq.~\eqref{eq:canonical} for a given system, we now show that there exists a procedure to approximately sample observables $O$ with a constant volume of support $k$ through only time evolution. First, we assume that the system Hamiltonian is composed of two noncommuting terms
\begin{equation}
    H=H_0+H_1,
\end{equation}
where an eigenstate $\ket{E^{\left(0\right)}}$ of $H_0$ is easy to prepare. Then, we consider the time evolution:
\begin{equation}
    \ket{\psi\left(t\right)}=\ce^{-\ci Ht}\ket{E^{\left(0\right)}};
\end{equation}
this procedure is a called a \emph{quench}. The long-time average of $\bra{\psi\left(t\right)}O\ket{\psi\left(t\right)}$ with Eq.~\eqref{eq:canonical} then gives:
\begin{equation}
%
%
    \overline{O}\equiv\lim\limits_{t\to\infty}\frac{1}{t}\int\limits_0^t\dd{t'}\bra{\psi\left(t'\right)}O\ket{\psi\left(t'\right)}\approx\frac{\tr\left(O\ce^{-\beta\left(E^{\left(0\right)}\right)H}\right)}{\tr\left(\ce^{-\beta\left(E^{\left(0\right)}\right)H}\right)}.
    \label{eq:time_average}
\end{equation}
This approximation is exact in the limit $n\to\infty$, given that energy fluctuations in $\ket{E^{\left(0\right)}}$ are small; $\beta\left(E^{\left(0\right)}\right)$ is an effective inverse temperature dictated by the initial state $\ket{E^{\left(0\right)}}$. In fact, it turns out that this equivalency is not only true in average, but also pointwise in time in the long time limit; in practice, however, the thermalization time is modest compared to the inverse norm of the Hamiltonian~\cite{DAlessio2016}. We give more details on the necessary assumptions and the degree of approximation in Appendix~\ref{sec:qbm_quench_therm}. Local thermalization after a quantum quench has been verified multiple times experimentally, including through the use of both superconducting qubits~\cite{neill2016ergodic} and ultracold atoms~\cite{Kaufman794}.

\section{Training QBMs Through ETH}\label{sec:thermal}

The quench procedure described in Sec.~\ref{sec:quench} prescribes a scheme with which to sample $k$-local observables in thermal distributions; as the observables that must be sampled are $1$- and $2$-local as given in Eq.~\eqref{eq:b_grad} and Eq.~\eqref{eq:w_grad}, these observables can be sampled using the given quench procedure~\footnote{Note, however, that the integrabiltiy restriction described in Sec.~\ref{sec:real_eth} means that this procedure has no equivalent in the classical regime. For a diagonal Hamiltonian, all observables that must be sampled in training commute with the Hamiltonian and are therefore conserved; therefore, ETH only implies thermalization up to to the local conservation of spin, and sampling diagonal observables in the time-evolved state will then be equivalent to sampling those observables in the initial state. See our brief discussion in Sec.~\ref{sec:real_eth} on generalized canonical ensembles for details.}. However, when using this scheme, $\beta$ in general is dependent on $\bm{\theta}$; though one could in principle control $\beta$ through coupling the QBM to a large bath, this would require many ancilla qubits. Instead, if we allow $\beta$ to become a function of $\bm{\theta}$, there are corrections to derivatives with respect to $\theta$ of the form:
\begin{equation}
    g_\theta\left(\bm{\theta}\right)=\pdv{\beta\left(\bm{\theta}\right)}{\theta}\sum\limits_{\bm{z_v}}p_{\textrm{data}}\left(\bm{z_v}\right)\left(\frac{\tr\left(H_{\bm{z_v}}\left(\bm{\theta}\right)\ce^{-\beta\left(\bm{\theta}\right)H_{\bm{z_v}}\left(\bm{\theta}\right)}\right)}{\tr\left(\ce^{-\beta\left(\bm{\theta}\right)H_{\bm{z_v}}\left(\bm{\theta}\right)}\right)}-\frac{\tr\left(H_{\textrm{QBM}}\left(\bm{\theta}\right)\ce^{-\beta\left(\bm{\theta}\right)H_{\textrm{QBM}}\left(\bm{\theta}\right)}\right)}{\tr\left(\ce^{-\beta\left(\bm{\theta}\right)H_{\textrm{QBM}}\left(\bm{\theta}\right)}\right)}\right).
    \label{eq:beta_correction}
\end{equation}
Thus, we need a method to estimate $\beta$ at various $\bm{\theta}$.

\begin{figure}[ht]
\resizebox{0.7\linewidth}{!}{
\begin{tikzpicture}
\node at (-3.75, 4) {\large QBM};
\draw (-5, -1) -- (-2.5, -2) node[anchor=north east,pos=0.65] {$w_{\upsilon\eta}$};
\draw (-5, -1) -- (-2.5, 0);
\draw (-5, -1) -- (-2.5, 2);
\draw (-5, 1) -- (-2.5, -2);
\draw (-5, 1) -- (-2.5, 0);
\draw (-5, 1) -- (-2.5, 2);
\draw[dashed] (-2.5, -2) -- (2.5, -2);
\draw[dashed] (-2.5, 2) -- (2.5, 2);
\draw (2.5, -2) -- (4.5, 0) node[anchor=north west,pos=0.5] {$w_{aa'}$};
\draw (2.5, -2) -- (2.5, 2);
\draw (2.5, 2) -- (4.5, 0);
\node[draw,circle,fill=blue!50,minimum size=30,inner sep=0] at (-5, -1) {$b_{\eta}$};
\node[draw,circle,fill=blue!50,minimum size=30,inner sep=0] at (-5, 1) {};
\draw[dotted,rounded corners] (-5.7, -2) rectangle (-4.3, 2) {};
\node[anchor=south, rotate=90] at (-6, 0) {Hidden Units};
\node[draw,circle,fill=green!50,minimum size=30,inner sep=0] at (-2.5, -2) {$b_{\upsilon}$};
\node[draw,circle,fill=green!50,minimum size=30,inner sep=0] at (-2.5, 0) {};
\node[draw,circle,fill=green!50,minimum size=30,inner sep=0] at (-2.5, 2) {};
\draw[dotted,rounded corners] (-3.3, -3) rectangle (-1.7, 3) {};
\node[anchor=south, rotate=270] at (-1.5, 0) {Visible Units};
\draw[dotted,rounded corners] (-6.7, -3.5) rectangle (-0.7, 3.5) {};
\node[draw,circle,fill=yellow!15,minimum size=30,inner sep=0] at (2.5, -2) {$b_a$};
\node[draw,circle,fill=yellow!15,minimum size=30,inner sep=0] at (2.5, 2) {};
\node[draw,circle,fill=yellow!15,minimum size=30,inner sep=0] at (4.5, 0) {};
\draw[dotted,rounded corners] (1, -3) rectangle (6, 3) {};
\node at (3.5, 3.5) {\large Thermometer};
\end{tikzpicture}
}
\caption{An example QBM/\allowbreak thermometer combination. The thermometer is weakly coupled to the QBM such that temperature measurements of the thermometer approximately agree with those of the entire system (see Sec.~\ref{sec:thermal}).}
\label{fig:thermo_qbm}
%
%
\end{figure}
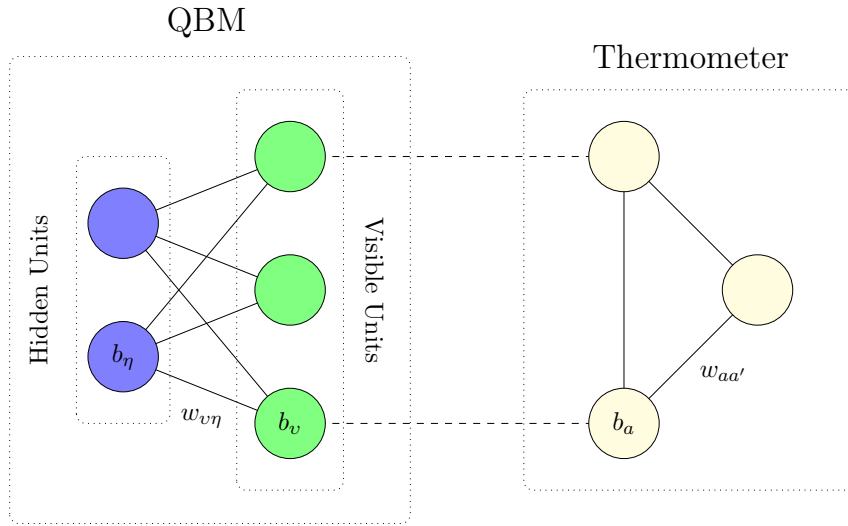

To do this, we couple the QBM system to an ancilla system of $\mathcal{O}\left(1\right)$ qubits we dub the \emph{thermometer system}. We fix a thermometer Hamiltonian of the same form as the QBM Hamiltonian with fixed parameters (see Appendix~\ref{sec:systems}).
%
%
%
%
%
%
%
%
We also fix an interaction Hamiltonian between the QBM and thermometer systems, given by:
\begin{equation}
    H_{\textrm{int}}=\sum\limits_{\upsilon',a}w_{\upsilon'a}\sigma_{\upsilon'}^z\sigma_a^z,
    \label{eq:int_ham}
\end{equation}
where $\upsilon'$ runs over a subset of the visible units of the QBM and $a$ over a subset of the thermometer system; see Fig.~\ref{fig:thermo_qbm} for a visual representation of the full system. We take $H_{\textrm{int}}$ to be sparse compared to the QBM Hamiltonian such that the measured local temperature of the thermometer system is approximately equal to the local temperature of the QBM system~\cite{Garrison2018}. We give explicit descriptions of the various QBM/\allowbreak thermometer pair systems we study in Appendix~\ref{sec:systems}.
%
%

Having set the QBM/\allowbreak thermometer Hamiltonian, we begin the sampling procedure in the pure state:
\begin{equation}
    \ket{\psi\left(t=0\right)}=\ket{+}^{\otimes n},
\end{equation}
which is an eigenstate of the off-diagonal part of the total Hamiltonian
\begin{equation}
    H=H_{\mathrm{QBM}}+H_{\textrm{therm}}+H_{\textrm{int}}.
\end{equation}
Then, we perform the time evolution:
\begin{equation}
    \ket{\psi\left(t\right)}=\ce^{-\ci Ht}\ket{\psi\left(0\right)}
\end{equation}
for $t\in\left\{T_i\right\}_i\equiv\mathcal{T}$. Under the conditions described in Sec.~\ref{sec:quench}, we have for all sites $i,j$ that:
%
%
\begin{align}
    \mathbb{E}_{t\sim\mathcal{T}}\left[\bra{\psi\left(t\right)}\sigma_i^z \ket{\psi\left(t\right)}\right]&\approx\frac{\tr\left(\sigma_i^z \ce^{-\beta H}\right)}{\tr\left(\ce^{-\beta H}\right)},\\
    \mathbb{E}_{t\sim\mathcal{T}}\left[\bra{\psi\left(t\right)}\sigma_i^z \sigma_j^z \ket{\psi\left(t\right)}\right]&\approx\frac{\tr\left(\sigma_i^z \sigma_j^z \ce^{-\beta H}\right)}{\tr\left(\ce^{-\beta H}\right)},\\
    \mathbb{E}_{t\sim\mathcal{T}}\left[\bra{\psi\left(t\right)}H_{\textrm{QBM}}\ket{\psi\left(t\right)}\right]&\approx\frac{\tr\left(H_{\textrm{QBM}}\ce^{-\beta H}\right)}{\tr\left(\ce^{-\beta H}\right)},\\
    \mathbb{E}_{t\sim\mathcal{T}}\left[\bra{\psi\left(t\right)}H_{\textrm{therm}}\ket{\psi\left(t\right)}\right]&\approx\frac{\tr\left(H_{\textrm{therm}}\ce^{-\beta H}\right)}{\tr\left(\ce^{-\beta H}\right)};\label{eq:therm_exp}
\end{align}
for details on the errors of these approximations, see Appendix~\ref{sec:qbm_quench_therm}. In principle, one may choose $\left\lvert\mathcal{T}\right\rvert=1$ (see Appendix~\ref{sec:qbm_quench_therm}), though choosing a larger $\left\lvert\mathcal{T}\right\rvert$ reduces the impact of fluctuations of observables away from their time average. In our training simulations in Sec.~\ref{sec:training}, we take $\left\lvert\mathcal{T}\right\rvert=2$.

To estimate $\beta\left(\bm{\theta}\right)$, we use the fact that Eq.~\eqref{eq:therm_exp} defines $\beta$ in the same sense as Eq.~\eqref{eq:e_beta_rel}. As $H_{\textrm{therm}}$ is known and has support on only $\mathcal{O}\left(1\right)$ qubits, one can numerically find $\beta$ by inverting Eq.~\eqref{eq:therm_exp} after estimating the expectation value of the thermometer Hamiltonian through sampling. Furthermore, given that $\left\lVert H_\text{int}\right\rVert$ is much smaller than $\left\lVert H_\text{QBM}\right\rVert$ and $\left\lVert H_\text{therm}\right\rVert$,
we expect that the measured inverse temperature of the thermometer is approximately that of the QBM~\cite{Garrison2018}. Thus, we can classically compute or approximately sample all terms in Eq.~\eqref{eq:b_grad}, Eq.~\eqref{eq:w_grad}, and Eq.~\eqref{eq:beta_correction}, and thus can train the parameters of the QBM efficiently.

Note, however, that drawn samples from the trained QBM/\allowbreak thermometer combination in general will not be able to recreate the many-body correlations of generic data distributions; this is because ETH only guarantees thermalization on small subsystems of the QBM. However, if these higher order correlations can be expressed in terms of lower order correlations, the QBM/\allowbreak thermometer combination can still potentially model the distribution. To see this, assume a fixed model for the data distribution over $n_v$ variables completely described by $m$ parameters. As there are $\mathcal{O}\left(n_v^k\right)$ components of the $k$th moment of the distribution, the data distribution model is completely determined by the first $k$ moments of the distribution~\footnote{This is assuming all components of the $k$th moment are independent; in general, if there are $\iota$ independent components of the $k$th moment, we must have that $m=O\left(n^{\iota\left(k\right)}\right)$.}, where
\begin{equation}
    m=\mathcal{O}\left(n_v^k\right).
\end{equation}
Thus, even though samples from the QBM/\allowbreak thermometer combination can only approximate the first $\mathcal{o}\left(n_v\right)$ moments of the true QBM sample distribution (see Appendix~\ref{sec:qbm_quench_therm}), this is sufficient for completely reconstructing classes of distributions completely parametrized by $\mathcal{o}\left(n_v^{n_v}\right)$ parameters through the method of moments~\cite{tchebycheff1890}. For instance, many classical data distributions---including distributions of images~\cite{1334543}---can be modeled as a mixture of Bernoulli distributions of the form:
%
%
\begin{equation}
    p\left(\bm{z_v}\right)=\frac{1}{m}\sum\limits_{i=1}^m p_{\textrm{Bernoulli}}\left(\bm{z_v};p_i,\bm{c}_i\right),
    \label{eq:mixed_bernoulli}
\end{equation}
where
\begin{equation}
    p_{\textrm{Bernoulli}}\left(\bm{z_v};p,\bm{c}\right)=p^{n_v-\left\lvert\frac{\bm{z_v}-\bm{c}}{2}\right\rvert}\left(1-p\right)^{\left\lvert\frac{\bm{z_v}-\bm{c}}{2}\right\rvert}
\end{equation}
is a Bernoulli distribution centered at $\bm{c}$ (here, $\left\lvert\bm{a}\right\rvert$ denotes the number of components equal to $-1$ of $\bm{a}$). As this distribution is completely described by only $2m$ parameters, for $m=\mathcal{o}\left(n_v^{n_v}\right)$ the parameters of the model (and thus the entire data distribution, assuming a fixed model) can be estimated by the QBM/\allowbreak thermometer combination. Furthermore, in practice, it seems numerically that sampling directly from the QBM/\allowbreak thermometer combination allows one to approximately sample from $p$ without explicitly reconstructing the model through the low order moments (see Sec.~\ref{sec:training}).

\section{Numerical Simulations}\label{sec:ns}

\subsection{Numerical Verification of Local Thermalization}\label{sec:ergodicity}

\begin{figure}[ht]
    \begin{center}
        \includegraphics[width=\linewidth]{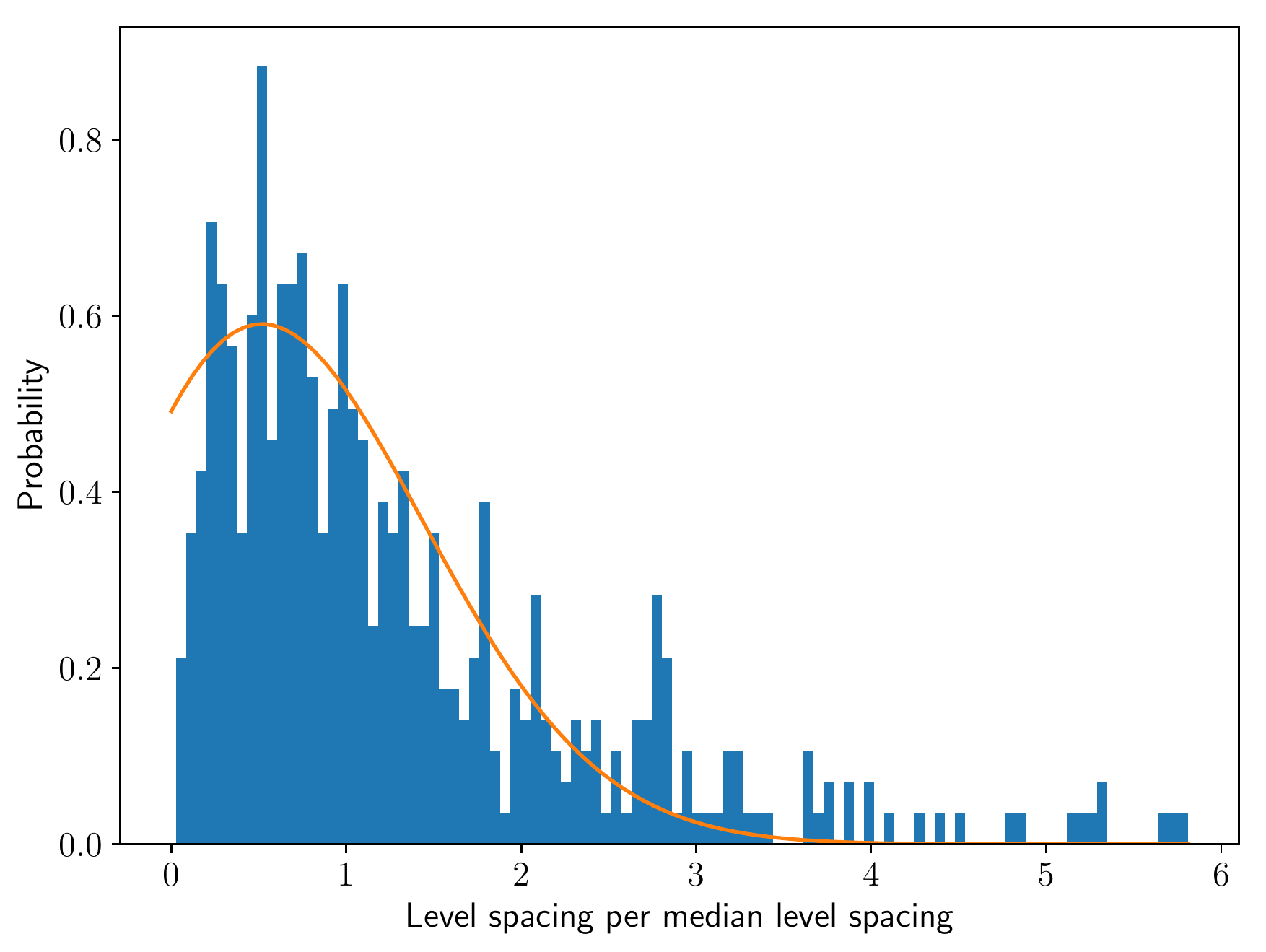}
        \caption{A typical fit of the Berry--Robnik distribution to the energy level spacing distribution of our trained QBM/\allowbreak thermometer combination. The trained model was a restricted transverse Ising model with $\frac{\overline{\varGamma}}{\sqrt{w_{\textrm{int}}^2}}=1$ (see Sec.~\ref{sec:ergodicity}), six visible units, one hidden unit, and two thermometer units.}
        %
        %
        \label{fig:berry_robnik_fit}
    \end{center}
\end{figure}

\begin{figure}[ht]
    \begin{center}
        \includegraphics[width=\linewidth]{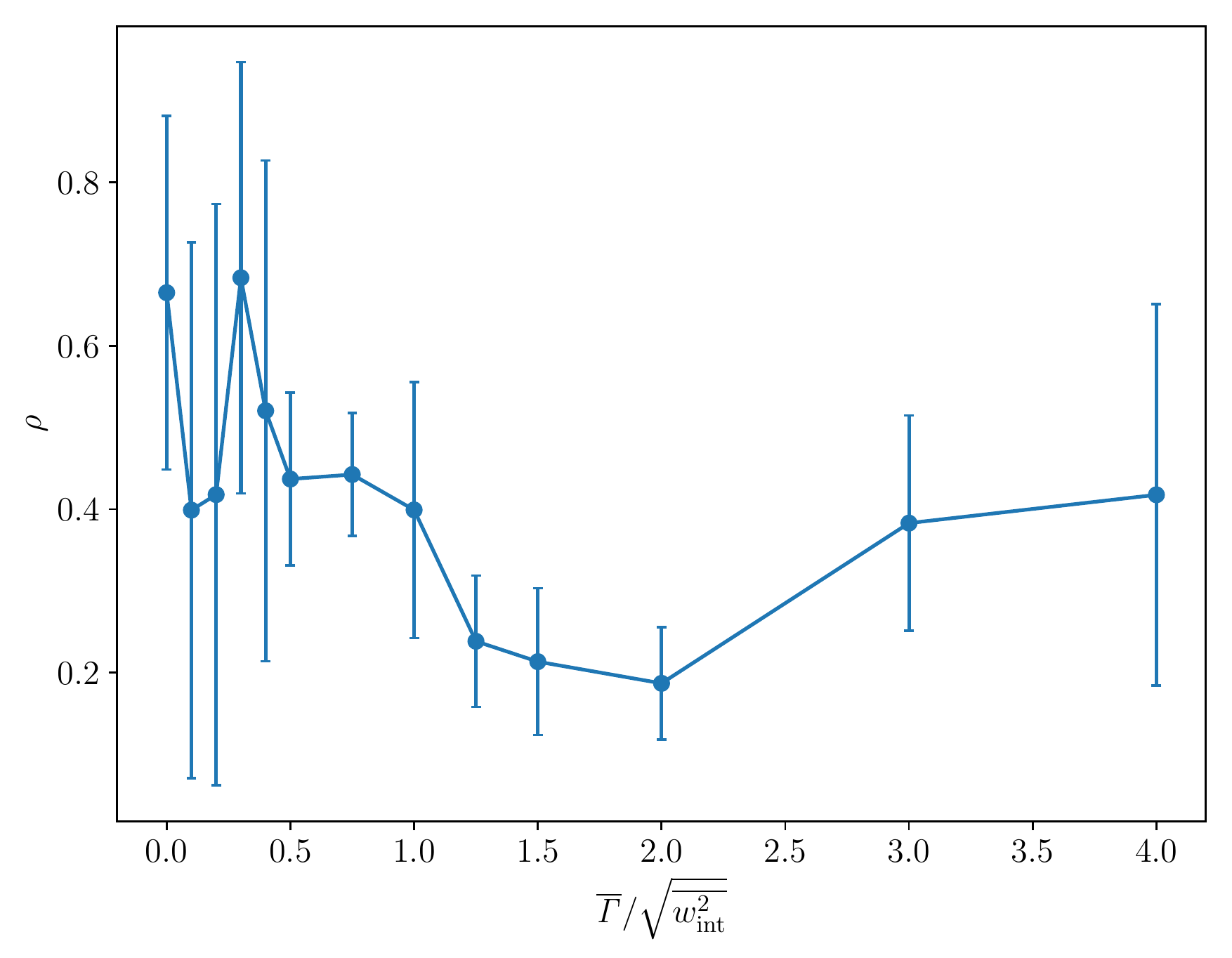}
        \caption{The Berry--Robnik interpolation parameter $\rho$ plotted as a function of the normalized mean single-site transverse field $\frac{\overline{\varGamma}}{\sqrt{w_{\textrm{int}}^2}}$ (see Sec.~\ref{sec:ergodicity}). The trained models were restricted transverse Ising models with six visible units, one hidden unit, and two thermometer units. Error bars denote one standard error over five instances.}
        %
        %
        \label{fig:berry_robnik_gamma}
    \end{center}
\end{figure}

We analyzed the distribution of energy level spacings to numerically verify the local thermalization of our quantum system. Apart from a few counterexamples, ideal ergodic systems obey a Wigner--Dyson distribution of energy level spacings, while ideal nonthermalizing systems obey a Poisson distribution in energy level spacings; in practice, systems interpolate between these two extremes~\cite{Berry_1984,DAlessio2016}. As an ansatz for this interpolating behavior we use the Berry--Robnik distribution~\cite{Berry_1984}, which is parametrized by some $\rho$ such that for $\rho=0$ the distribution is identical to a Wigner--Dyson distribution and for $\rho=1$ the distribution is identical to a Poisson distribution. In all cases, we normalize our empirical distributions by the median energy level spacing. Fig.~\ref{fig:berry_robnik_fit} is a typical fit of the Berry--Robnik distribution to the energy level spacing distribution of our trained QBM/\allowbreak thermometer combination. Fig.~\ref{fig:berry_robnik_gamma} shows fits of the Berry--Robnik interpolation parameter $\rho$ for various values of the mean single-site transverse field $\overline{\varGamma}$ for a restricted transverse Ising model, normalized by the root mean square of the interaction weights between the QBM and thermometer $\sqrt{w_{\textrm{int}}^2}$.
%
%
We see for $1\lesssim\frac{\overline{\varGamma}}{\sqrt{w_{\textrm{int}}^2}}\lesssim 3$ that our QBM/\allowbreak thermometer combination has an energy level spacing distribution consistent with that of a chaotic system. Furthermore, we see as $\overline{\varGamma}\to 0$ that this is no longer true; this is due to the restricted transverse Ising model reducing to the classical restricted Ising model in this limit, which conserves local spin and therefore is not expected to thermalize to a canonical distribution locally.

\begin{figure*}[ht]
    \begin{center}
        \begin{subfloat}[Semi-restricted Transverse Ising Model]{\resizebox{0.45\linewidth}{!}{\includegraphics[width=\linewidth]{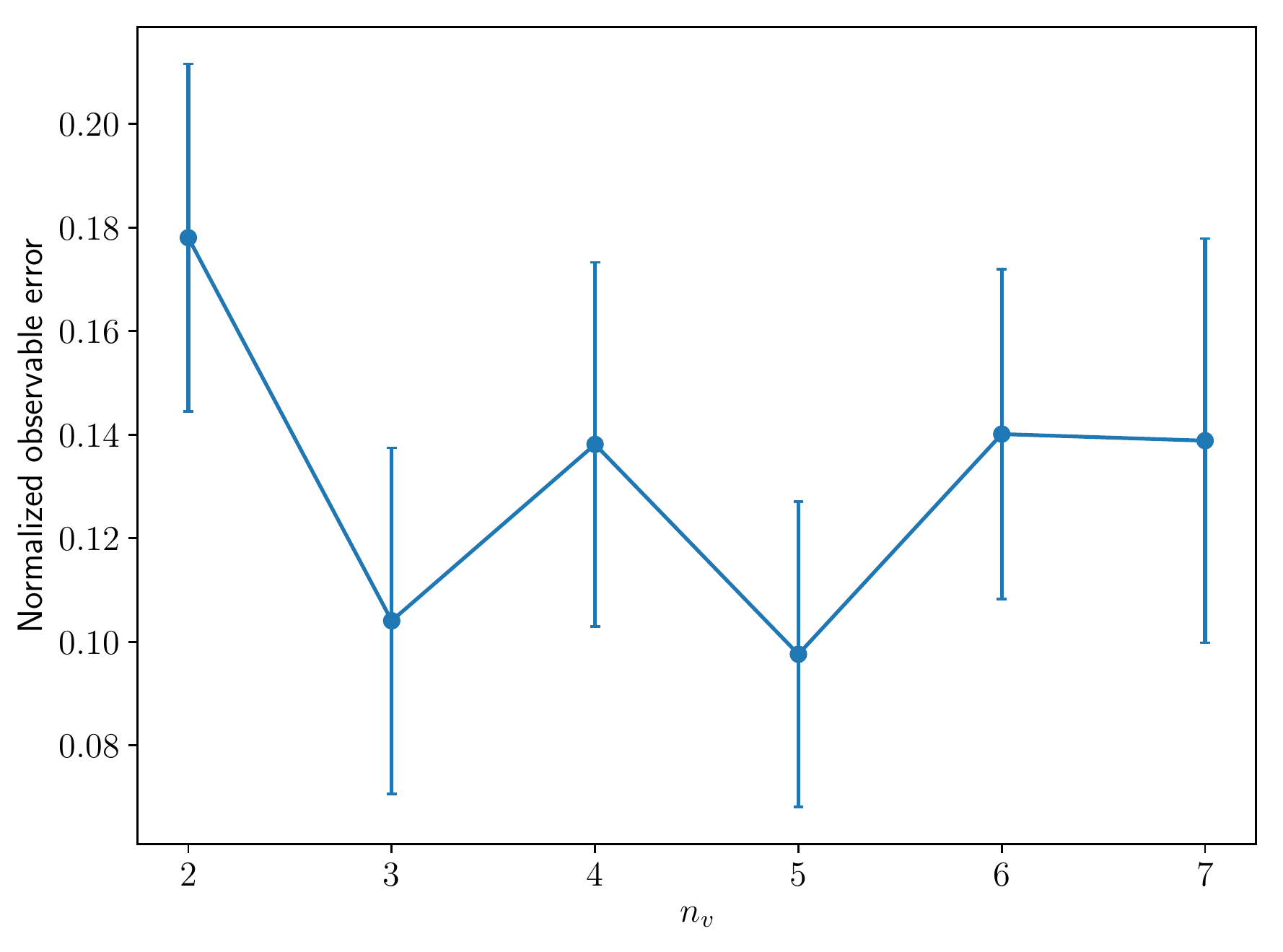}}}\end{subfloat}
        \begin{subfloat}[Restricted Transverse Ising Model]{\resizebox{0.45\linewidth}{!}{\includegraphics[width=\linewidth]{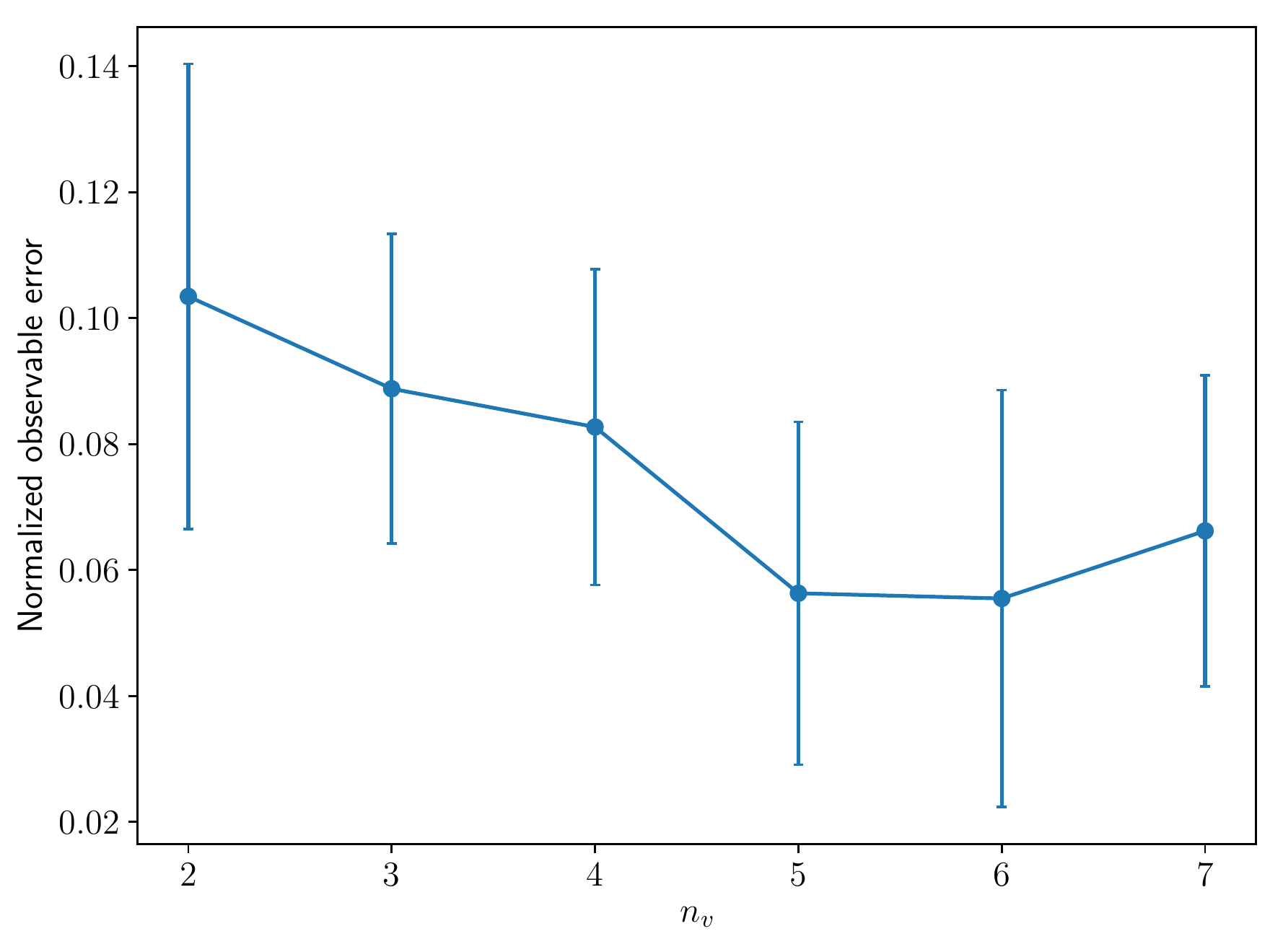}}}\end{subfloat}
        \begin{subfloat}[Restricted XX Model]{\resizebox{0.45\linewidth}{!}{\includegraphics[width=\linewidth]{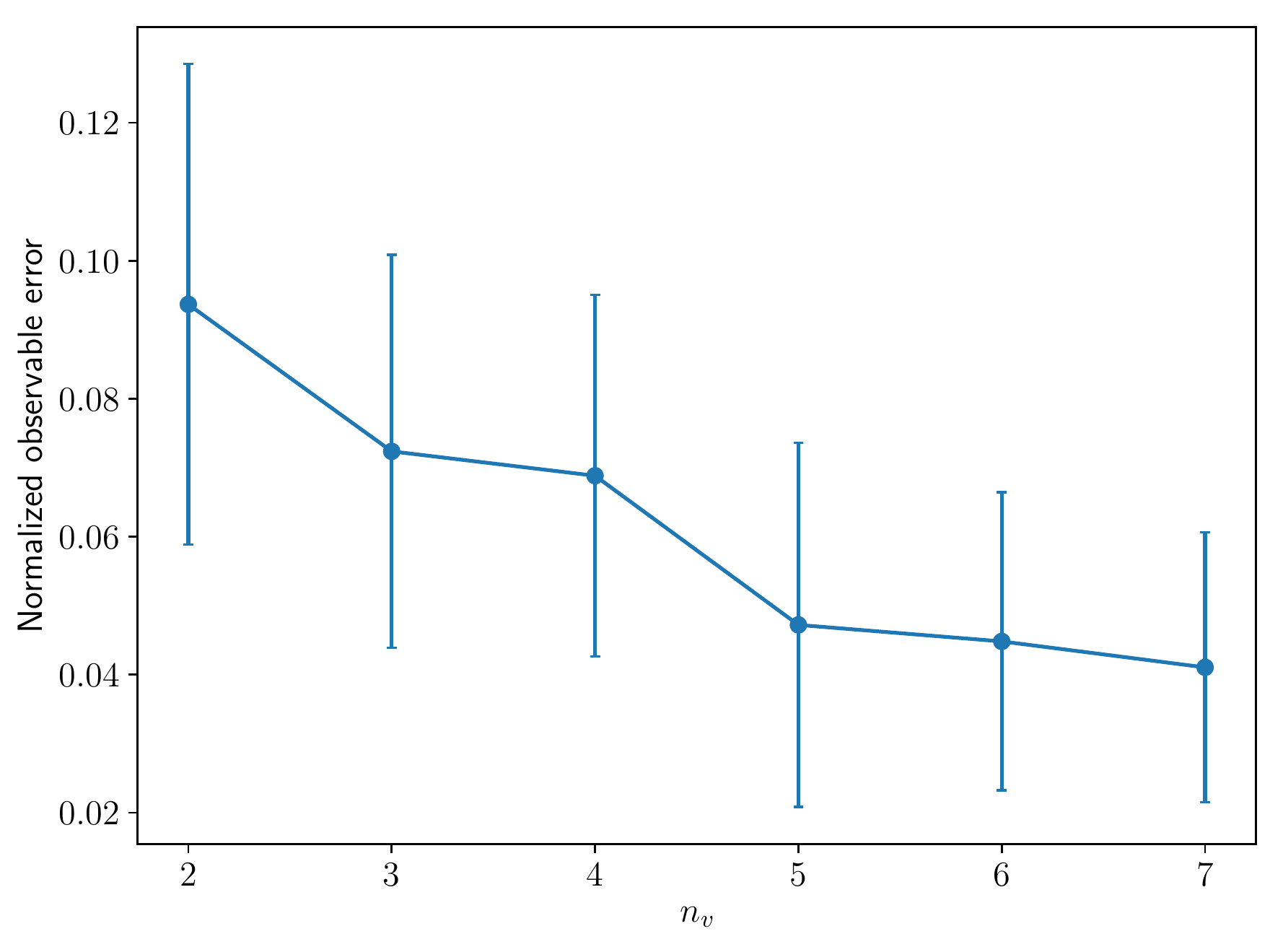}}}\end{subfloat}
        \caption{The median error in gradient observable using our QBM/\allowbreak thermometer scheme for multiple Hamiltonian models as a function of the number of visible units of the system (see Sec.~\ref{sec:ergodicity}). The lack of a convergence in the thermodynamic limit for the semi-restricted transverse Ising model is most likely due to a nonvanishing variance of the energy expectation value of the system in the thermodynamic limit (see Appendix~\ref{sec:qbm_quench_therm}). The studied models each had one hidden unit and two thermometer units; for greater detail on the studied systems, see Appendix~\ref{sec:systems}. Error bars denote one standard error over five instances and over all gradient observables.}
        \label{fig:local_therm_size}
    \end{center}
\end{figure*}

\begin{figure*}[ht]
    \begin{center}
        \begin{subfloat}[Semi-restricted Transverse Ising Model]{\resizebox{0.45\linewidth}{!}{\includegraphics[width=\linewidth]{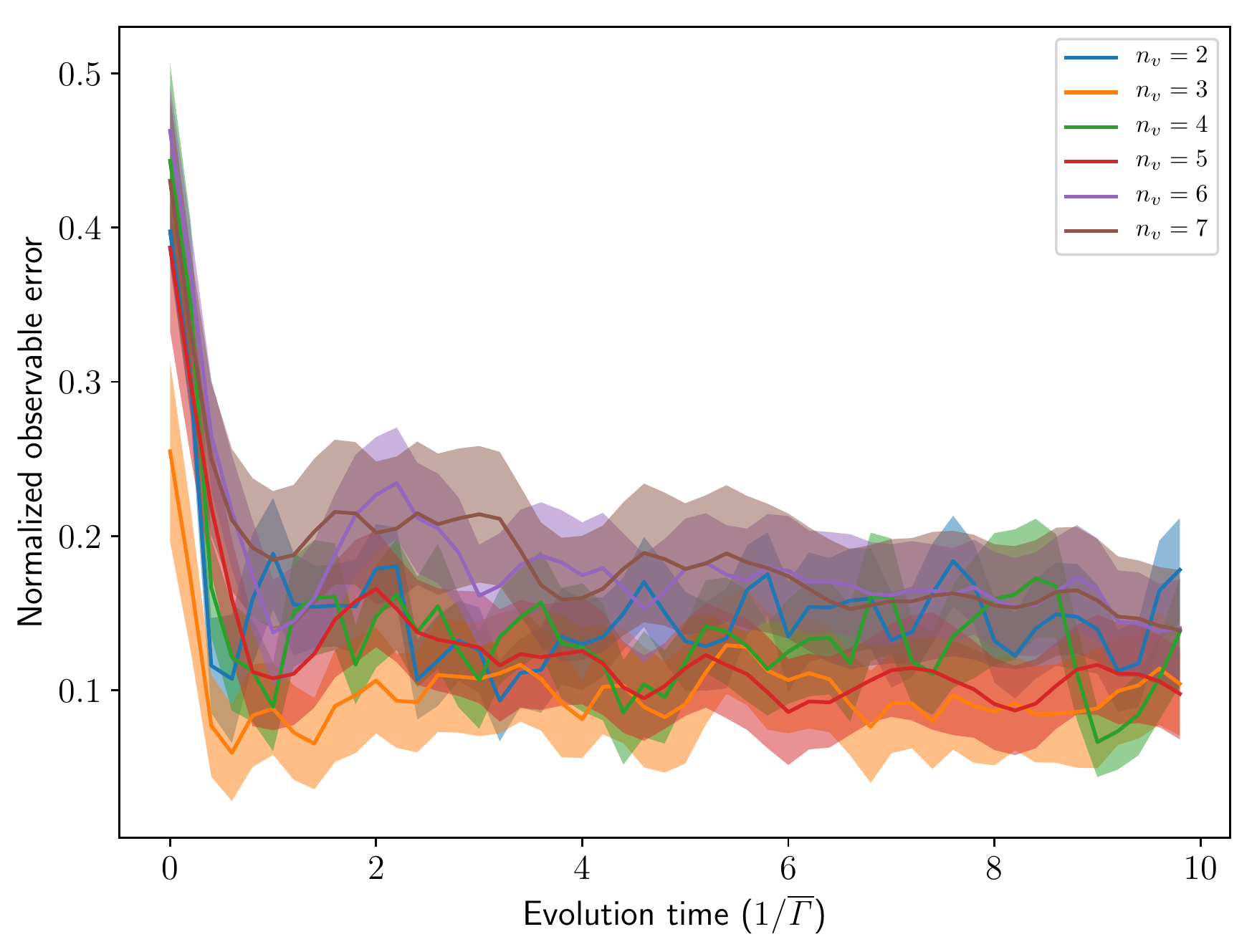}}}\end{subfloat}
        \begin{subfloat}[Restricted Transverse Ising Model]{\resizebox{0.45\linewidth}{!}{\includegraphics[width=\linewidth]{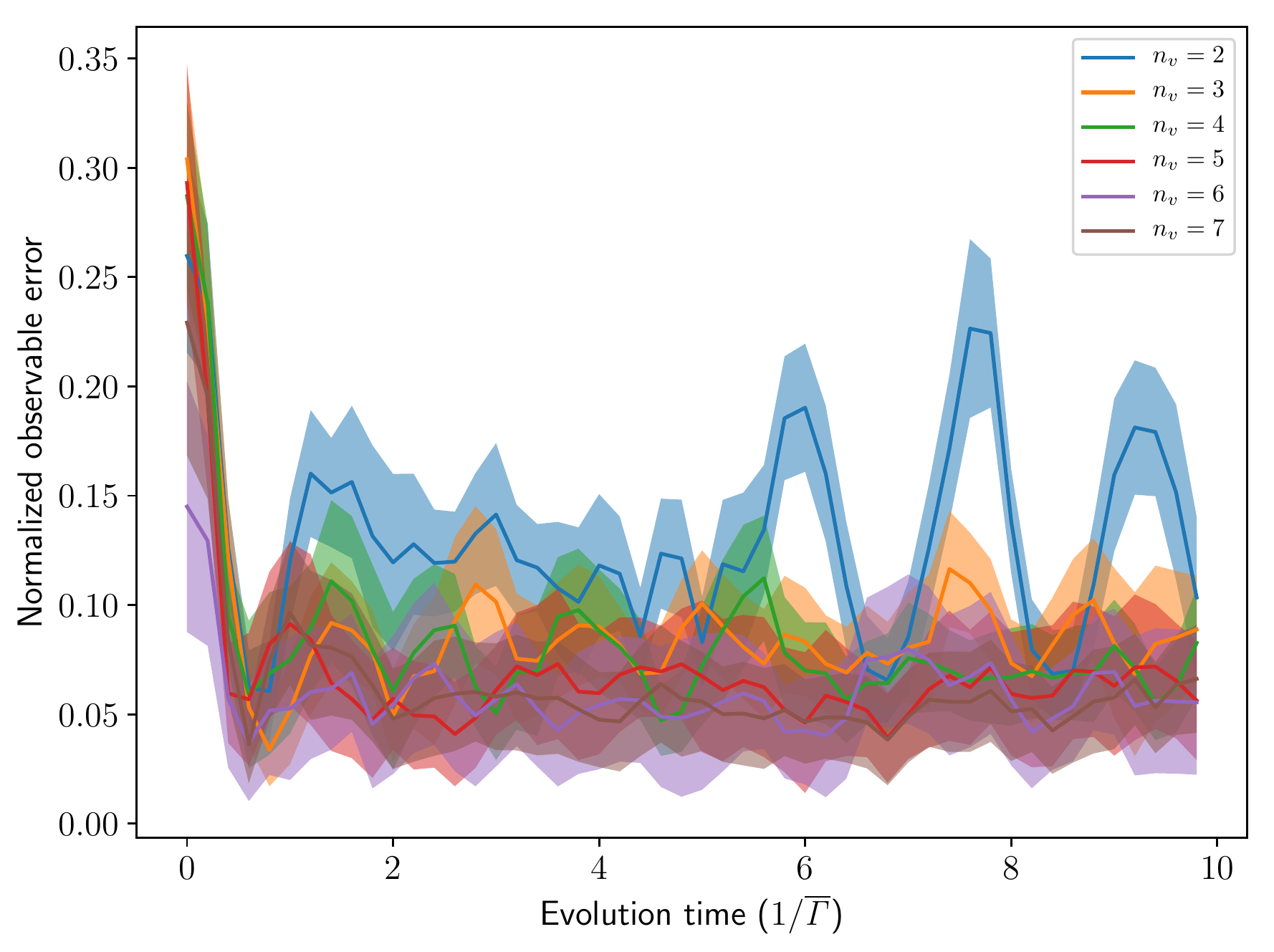}}}\end{subfloat}
        \begin{subfloat}[Restricted XX Model]{\resizebox{0.45\linewidth}{!}{\includegraphics[width=\linewidth]{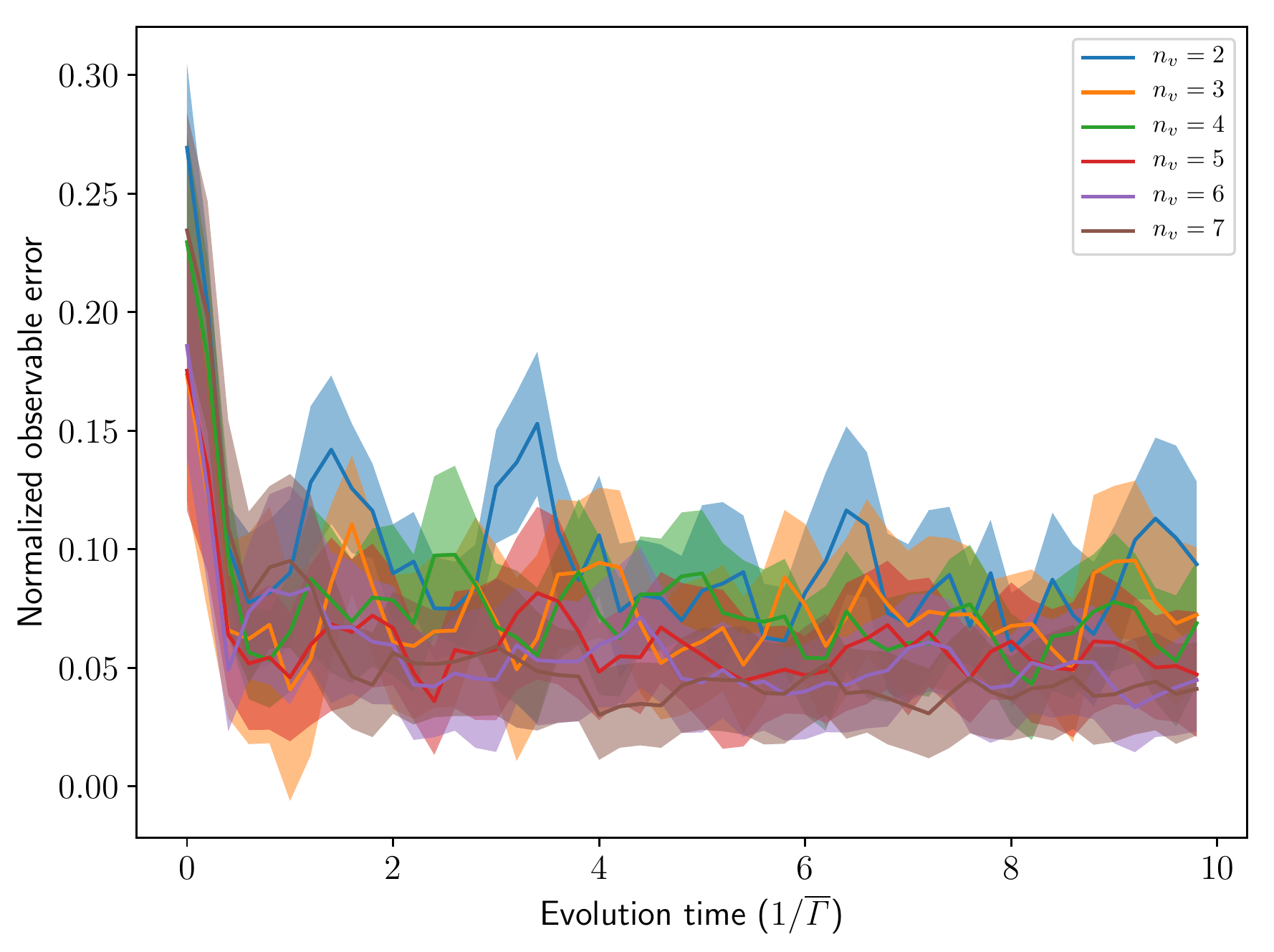}}}\end{subfloat}
        \caption{The median error in gradient observable using our QBM/\allowbreak thermometer scheme for multiple Hamiltonian models as a function of the quench evolution time (see Sec.~\ref{sec:ergodicity}). The studied models each had one hidden unit and two thermometer units; for greater detail on the studied systems, see Appendix~\ref{sec:systems}. Shading denotes one standard error over five instances.}
        \label{fig:local_therm_time}
    \end{center}
\end{figure*}

\begin{figure*}[ht]
    \begin{center}
        \begin{subfloat}[Semi-restricted Transverse Ising Model]{\resizebox{0.45\linewidth}{!}{\includegraphics[width=\linewidth]{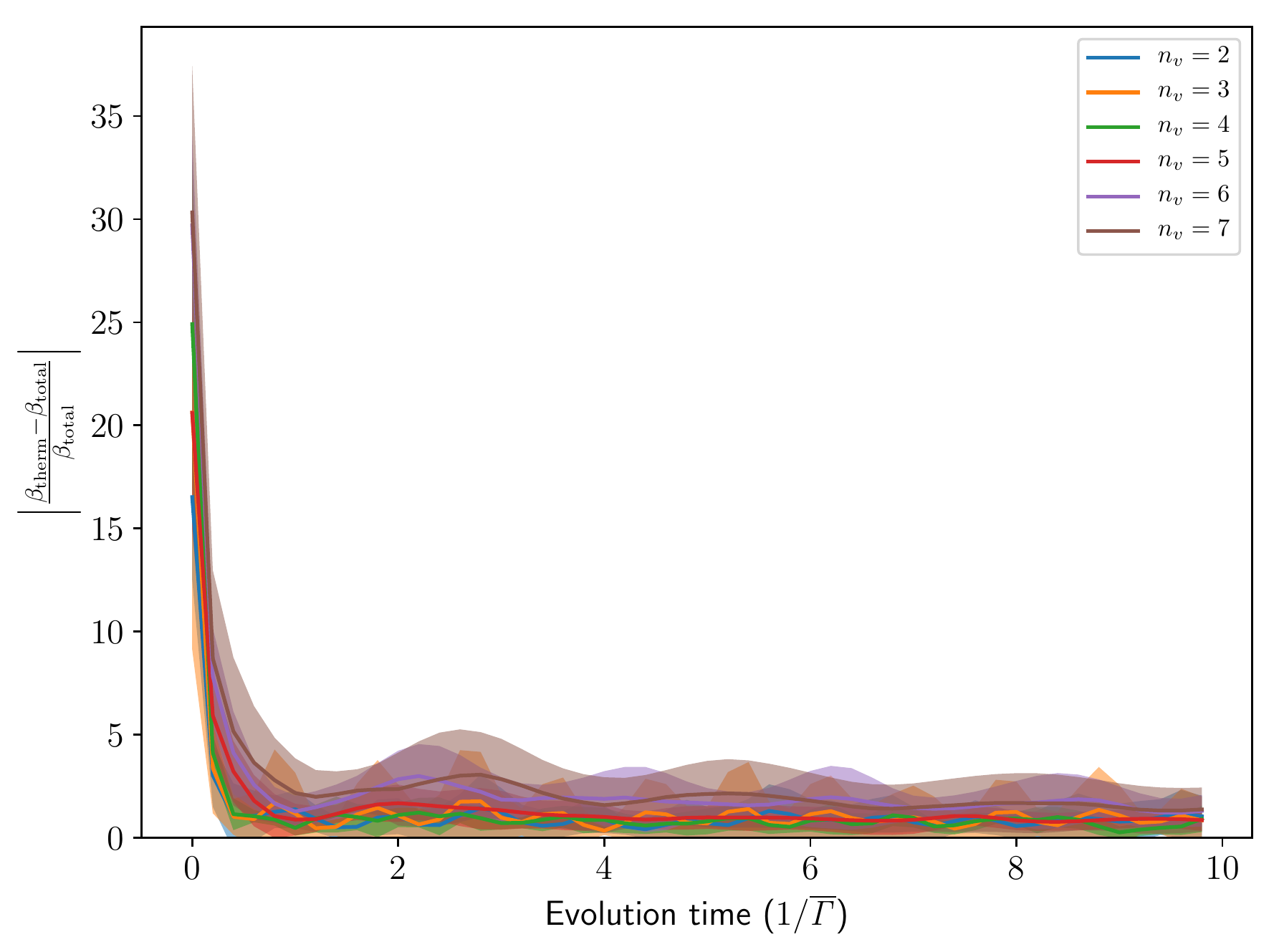}}}\end{subfloat}
        \begin{subfloat}[Restricted Transverse Ising Model]{\resizebox{0.45\linewidth}{!}{\includegraphics[width=\linewidth]{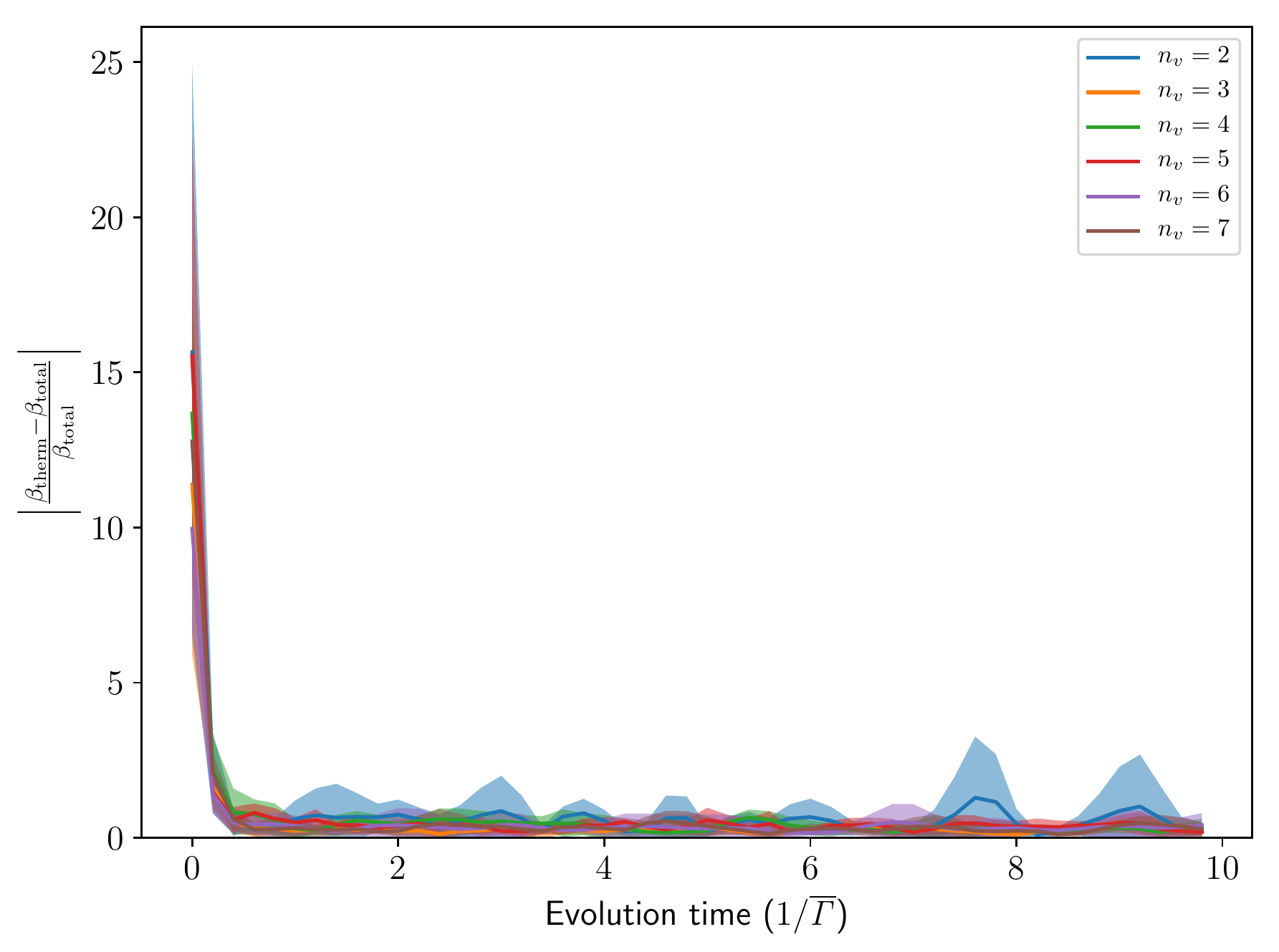}}}\end{subfloat}
        \begin{subfloat}[Restricted XX Model]{\resizebox{0.45\linewidth}{!}{\includegraphics[width=\linewidth]{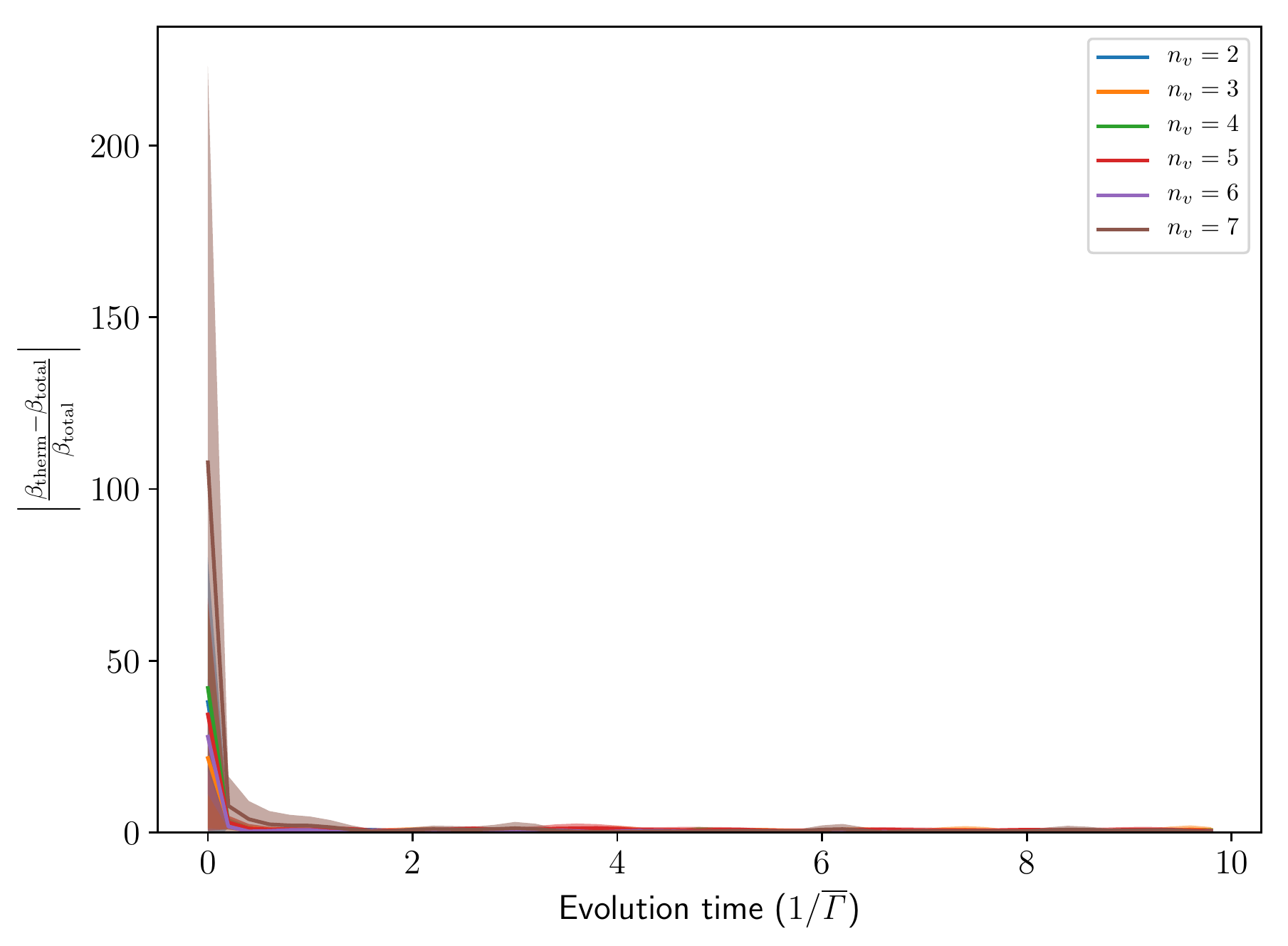}}}\end{subfloat}
        \caption{The mean normalized error in the inverse temperature of the thermometer for multiple Hamiltonian models as a function of the quench evolution time (see Sec.~\ref{sec:ergodicity}). The studied models each had one hidden unit and two thermometer units; for greater detail on the studied systems, see Appendix~\ref{sec:systems}. Shading denotes one standard error over five instances.}
        \label{fig:therm_equil_time}
    \end{center}
\end{figure*}

We also verified the local thermalization of gradient observables using our quenching procedure by measuring the median difference in gradient observable expectation value, normalized over the maximum possible error of the observable (namely, for Pauli matrices, two). We performed our numerical experiments on constructed models with parameters given in Appendix~\ref{sec:systems}, chosen to approximate typical values of the parameters of Hamiltonians following our training procedure. Fig.~\ref{fig:local_therm_size} demonstrates the local thermalization of gradient observables (i.e. the observables of Eq.~\eqref{eq:grad}) for various numbers of visible units. We see that the normalized error in observable expectation value decreases for larger system size for the restricted architectures, but this behavior is less obvious for the semi-restricted architecture; we believe this is due to the largeness of energy fluctuations of the system, as described in Appendix~\ref{sec:qbm_quench_therm}. Furthermore, the time needed for the system to thermalize is on the order of the reciprocal mean single-site transverse field $1/\overline{\varGamma}$, as demonstrated in Fig.~\ref{fig:local_therm_time}. The thermalization time seems independent of the system size, which is comparable with similar numerical experiments~\cite{DAlessio2016}. Finally, the equilibration of the thermometer with the system occurs extremely quickly, with the measured inverse temperature $\beta_{\textrm{therm}}$ as predicted by the thermometer Hamiltonian quickly converging to the measured inverse temperature $\beta_{\textrm{total}}$ as predicted by using the full system Hamiltonian, as demonstrated in Fig.~\ref{fig:therm_equil_time}.

%

%

\subsection{Numerical Verification of Training}\label{sec:training}

To analyze the efficacy of training quantum Boltzmann machines using our methods, we numerically simulated training QBM/\allowbreak thermometer combinations on mixtures of Bernoulli distributions of the form given in Eq.~\eqref{eq:mixed_bernoulli}. We took $m=8$, $p_i=0.9$, uniformly random $\bm{c}_i$, and $n_v\in\{2,3,4,5,6,7,8\}$ for a variety of models described in Appendix~\ref{sec:systems}. The exact training parameters we used are described in Appendix~\ref{sec:training_procedure}.

To evaluate the performance of our training procedure, we sampled from both the data distribution and the trained models to estimate the Kullback--Leibler (KL) divergence~\cite{kullback1951} of the data distribution with the model distribution
\begin{equation}
    D_{\textrm{KL}}\left(\left.\left.p_{\textrm{data}}\right|\right| p_{\textrm{model}}\right)=-\sum\limits_{\bm{z_v}}p_{\textrm{data}}\left(\bm{z_v}\right)\ln\left(\frac{p_{\textrm{model}}\left(\bm{z_v}\right)}{p_{\textrm{data}}\left(\bm{z_v}\right)}\right)
\end{equation}
and the Akaike information criterion (AIC)~\cite{Akaike1974}
\begin{equation}
    \operatorname{AIC}\left(\bm{\theta}\right)=2\left(\left\lvert\bm{\theta}_{\textrm{trainable}}\right\rvert+\mathcal{L}\left(\bm{\theta}\right)\right),
\end{equation}
where $\bm{\theta}_{\textrm{trainable}}$ are the trainable parameters of the model and $\mathcal{L}$ is the negative log-likelihood given in Eq.~\eqref{eq:exact_ll}. The KL divergence---though not a true metric---is a premetric between probability distributions. The AIC is similar, though also penalizes the number of trainable parameters in the model. We also implemented an amplitude damping channel with single-qubit Kraus operators:
\begin{align}
    E_1^{\left(a\right)}\left(t\right)=\begin{pmatrix}
    1 & 0\\
    0 & \ce^{-\frac{t}{2T_1}}
    \end{pmatrix},\\
    E_2^{\left(a\right)}\left(t\right)=\begin{pmatrix}
    0 & \sqrt{1-\ce^{-\frac{t}{T_\phi}}}\\
    0 & 0
    \end{pmatrix},
\end{align}
and a dephasing channel with single-qubit Kraus operators:
\begin{align}
    E_1^{\left(d\right)}\left(t\right)=\begin{pmatrix}
    1 & 0\\
    0 & \ce^{-\frac{t}{T_\phi}}
    \end{pmatrix},\\
    E_2^{\left(d\right)}\left(t\right)=\begin{pmatrix}
    0 & 0\\
    0 & \sqrt{1-\ce^{-\frac{2t}{T_\phi}}}
    \end{pmatrix},
\end{align}
where $t$ is randomly sampled from the distribution of times used to perform the Hamiltonian evolution in the quench procedure (see Sec.~\ref{sec:quench}). We additionally estimated the effects of sampling noise by including Gaussian noise on the sampling of each operator (see Appendix~\ref{sec:training_procedure}).

\begin{figure*}[ht]
    \begin{center}
        \begin{subfloat}[Semi-restricted Transverse Ising Model]{\resizebox{0.45\linewidth}{!}{\includegraphics[width=\linewidth]{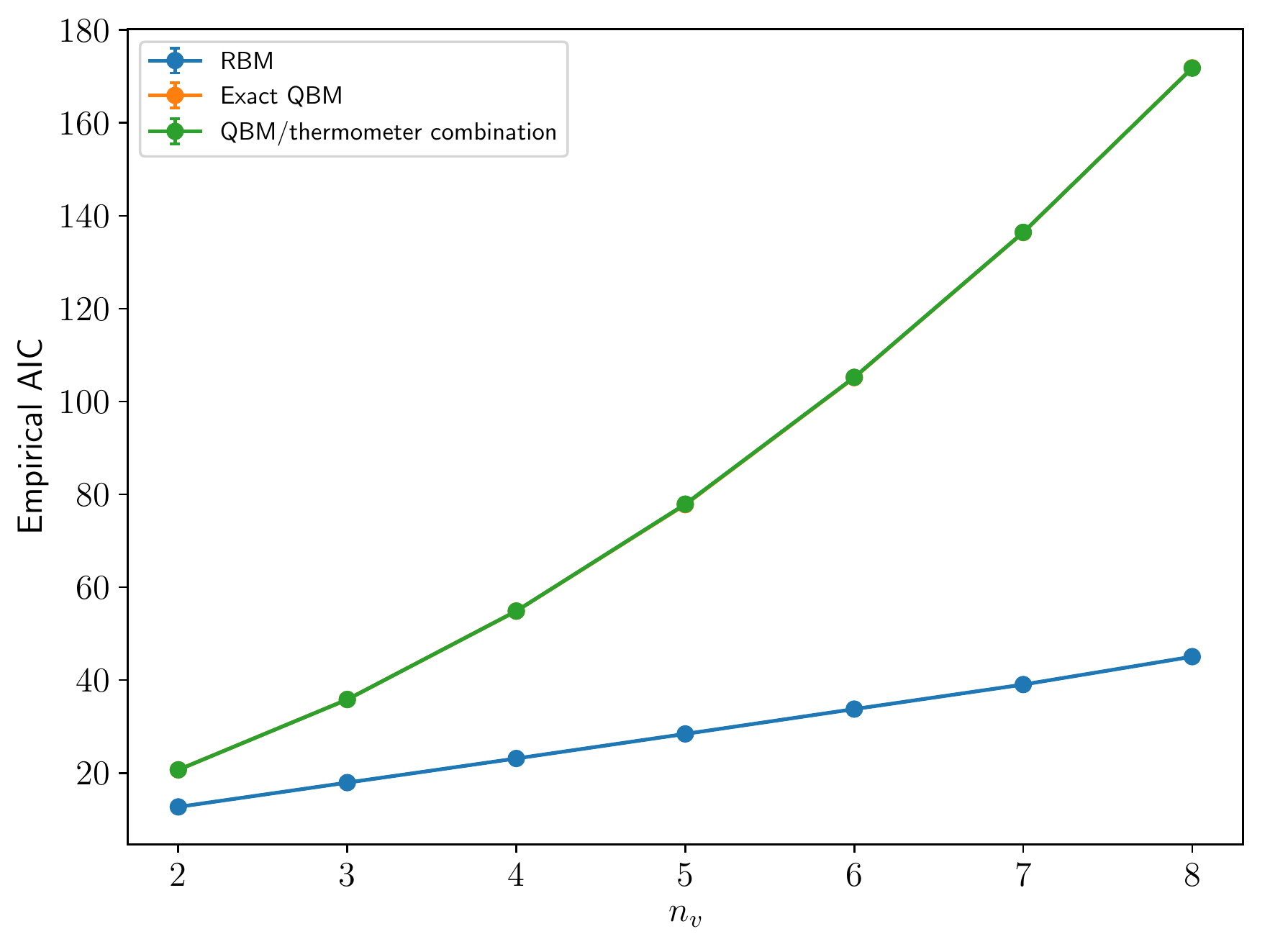}}}\end{subfloat}
        \begin{subfloat}[Restricted Transverse Ising Model]{\resizebox{0.45\linewidth}{!}{\includegraphics[width=\linewidth]{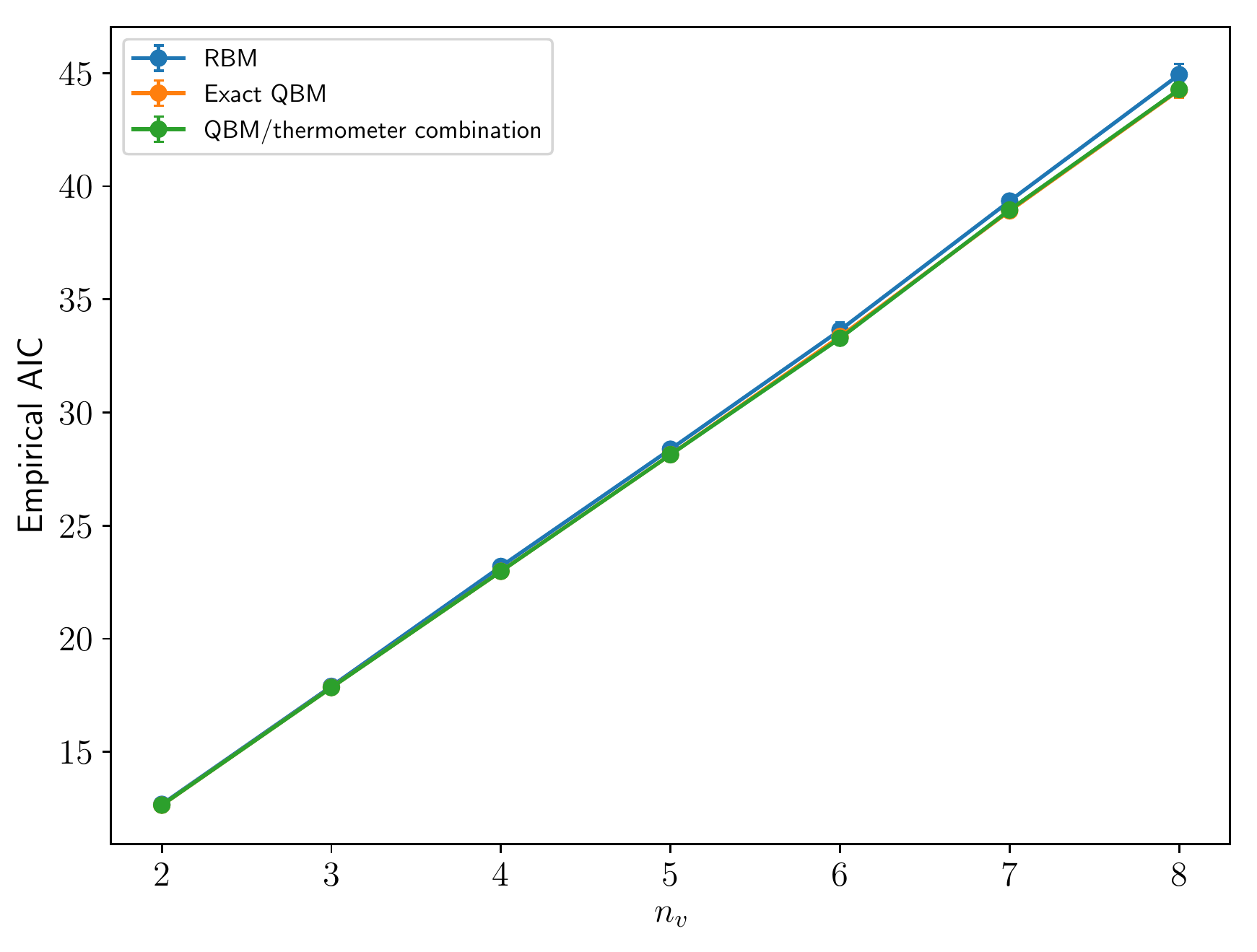}}}\end{subfloat}
        \begin{subfloat}[Restricted XX Model]{\resizebox{0.45\linewidth}{!}{\includegraphics[width=\linewidth]{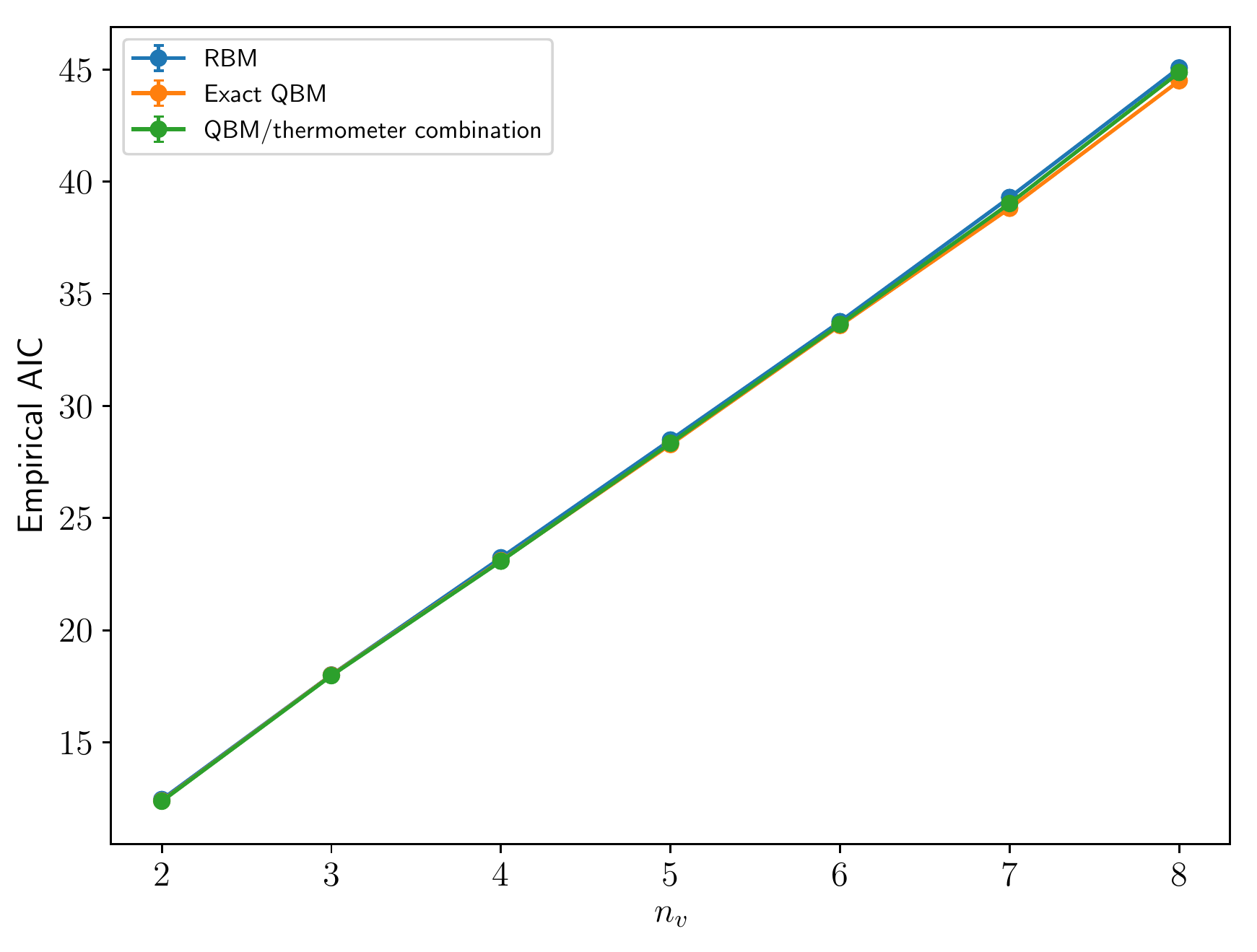}}}\end{subfloat}
        \caption{The training performance of our combined QBM/\allowbreak thermometer scheme on $n_v$-dimensional Bernoulli mixture models for multiple Hamiltonian models, also taking into account the number of trained parameters. A lower AIC with the data distribution corresponds with better performance, with a linear penalty applied for the number of trained parameters. In terms of AIC, the semi-restricted transverse Ising model for even an exact QBM is outperformed by an RBM due to the many visible-visible layer couplings in this model. The studied models are described in greater detail in Appendix~\ref{sec:systems}. Error bars denote one standard error over five instances.}
        %
        %
        \label{fig:aics}
    \end{center}
\end{figure*}

\begin{figure}[ht]
    \begin{center}
        \includegraphics[width=\linewidth]{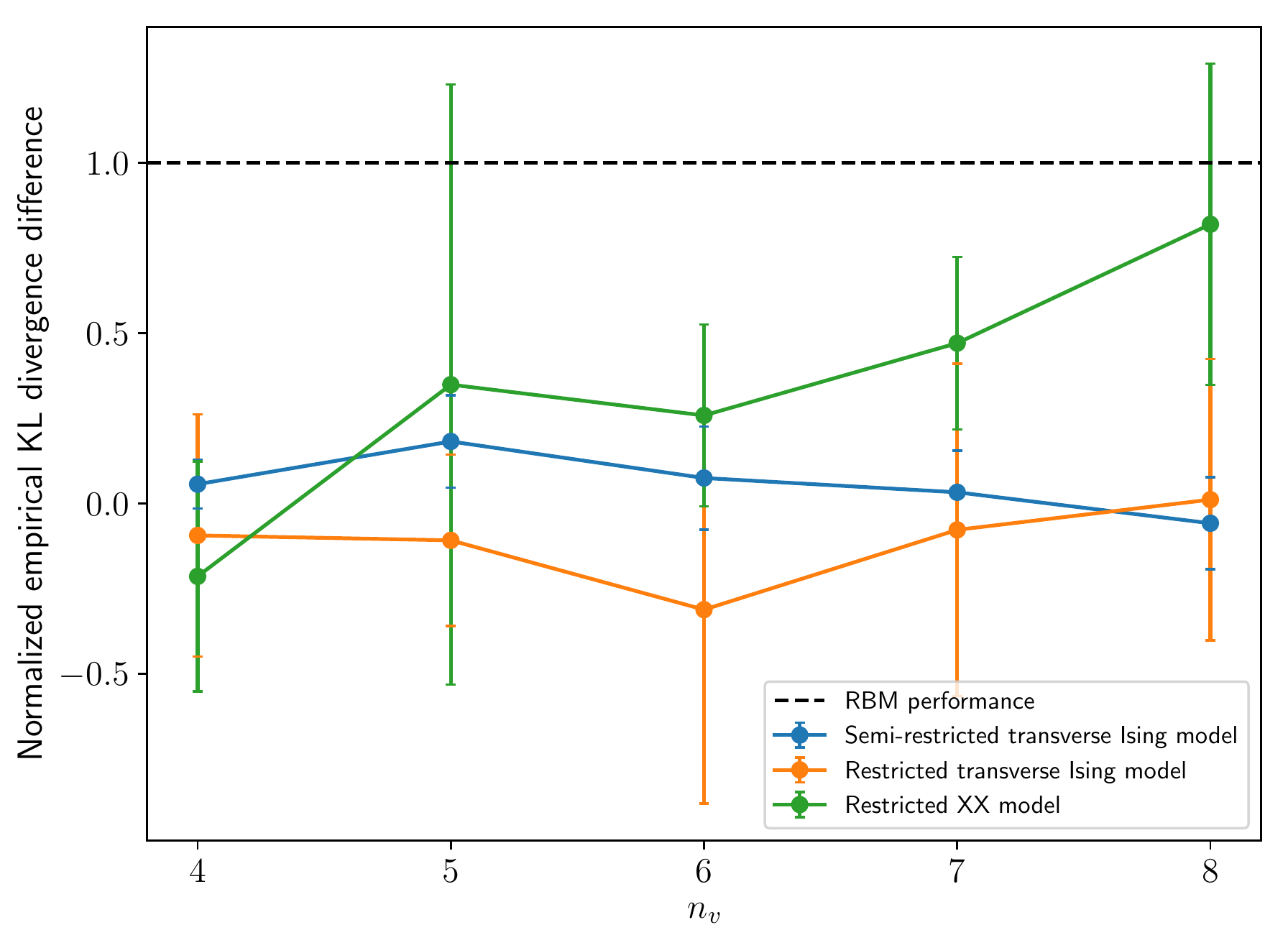}
        \caption{The ratio of the difference between the minimum achieved KL divergence for a QBM/\allowbreak thermometer combination and an exact QBM, and that of an RBM and an exact QBM is plotted for various mixed Bernoulli distribution dimensions and for multiple Hamiltonian models. Training performance below the dashed line demonstrates that the QBM/\allowbreak thermometer combination outperforms the classical RBM. Error bars denote one standard error over five instances.}
        %
        %
        \label{fig:kl_div_diffs}
    \end{center}
\end{figure}

In Fig.~\ref{fig:kl_divs}, we plot the minimum reached KL divergence during training as a function of the dimensionality of the data distribution $n_v$; similarly, in Fig.~\ref{fig:aics}, we plot the minimum reached AIC. To summarize the performance of our QBM/\allowbreak thermometer combination compared with an exact QBM (i.e. one with oracle access to quantum thermal states), in Fig.~\ref{fig:kl_div_diffs} we plot the ratio between the difference in optimized KL divergences between a QBM/\allowbreak thermometer combination and an ideal QBM, with the difference between an RBM and an ideal QBM.
%
%

\begin{figure*}[ht]
    \begin{center}
        \begin{subfloat}[Semi-restricted Transverse Ising Model]{\resizebox{0.45\linewidth}{!}{\includegraphics[width=\linewidth]{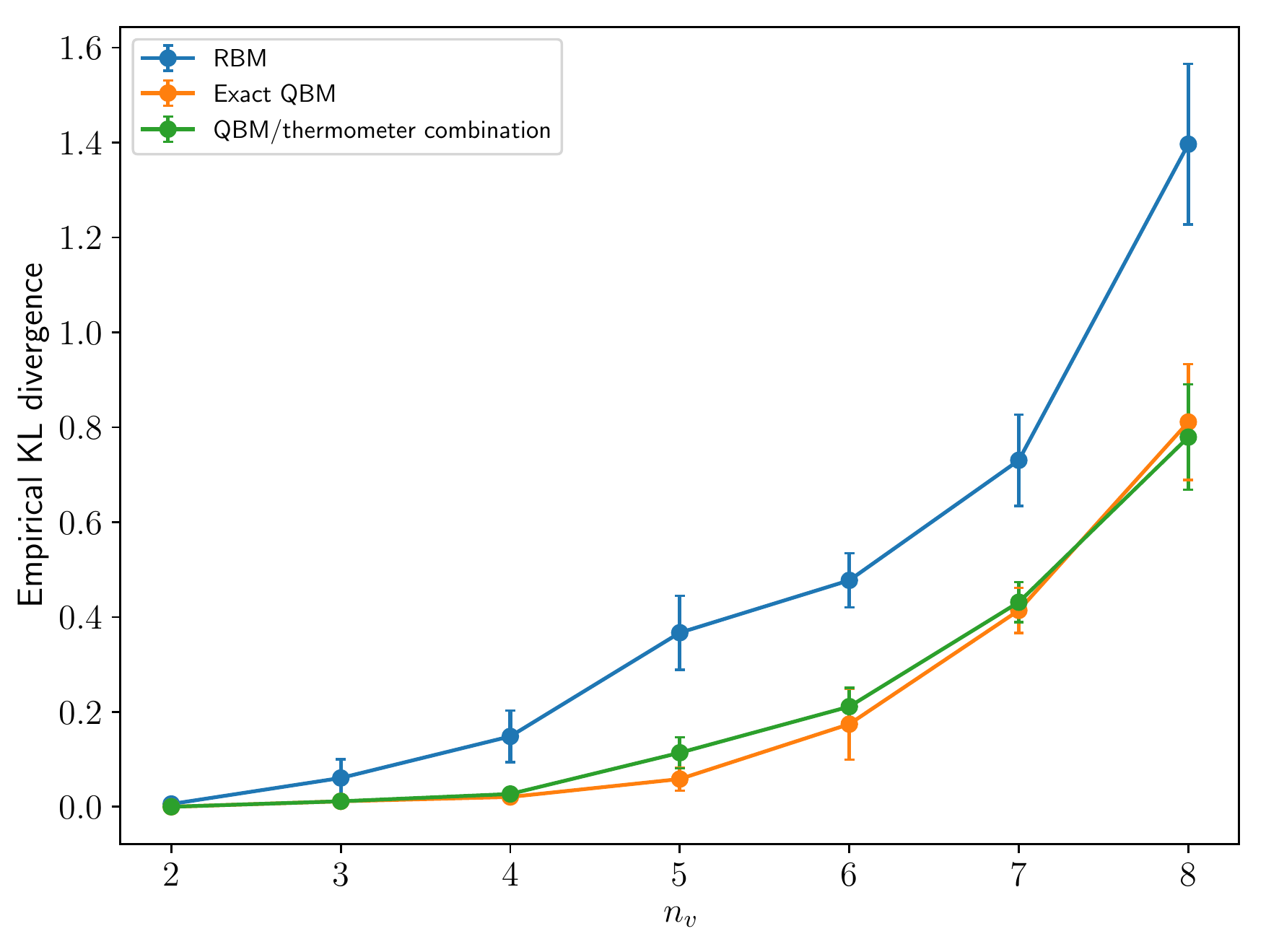}}}\end{subfloat}
        \begin{subfloat}[Restricted Transverse Ising Model]{\resizebox{0.45\linewidth}{!}{\includegraphics[width=\linewidth]{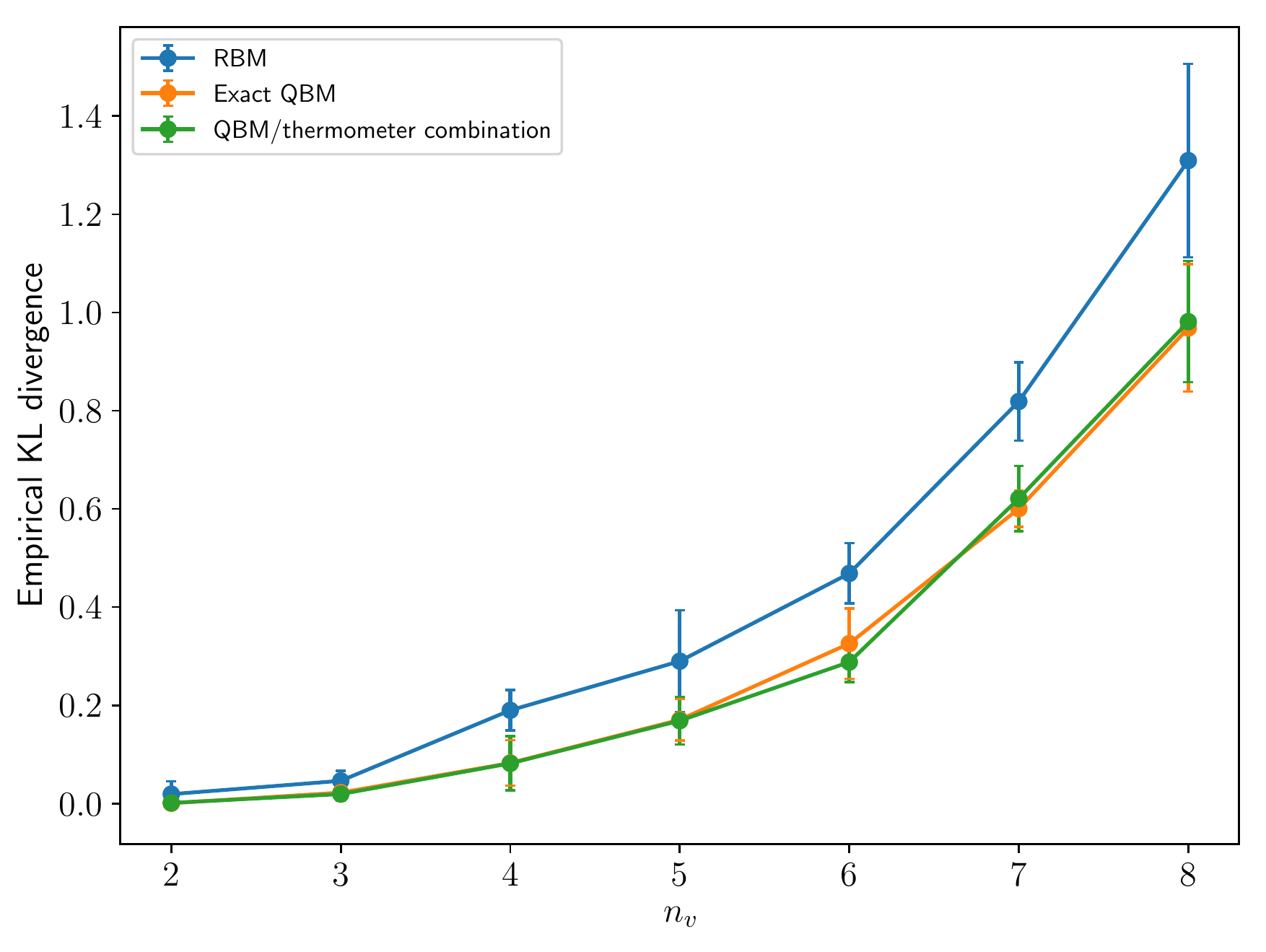}}}\end{subfloat}
        \begin{subfloat}[Restricted XX Model]{\resizebox{0.45\linewidth}{!}{\includegraphics[width=\linewidth]{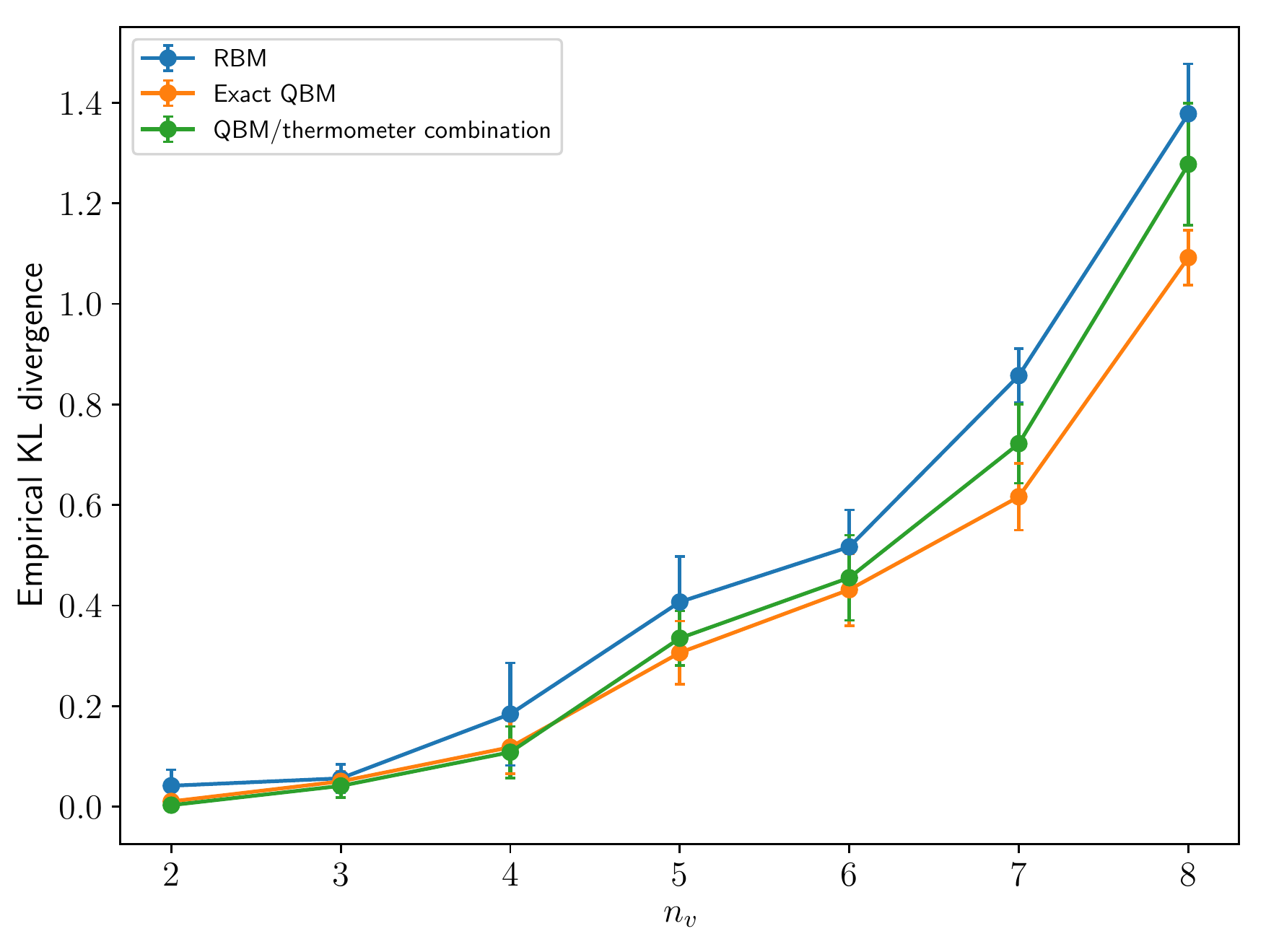}}}\end{subfloat}
        \caption{The training performance of our combined QBM/\allowbreak thermometer scheme on $n_v$-dimensional Bernoulli mixture models for multiple Hamiltonian models. A lower KL divergence with the data distribution corresponds with better performance. For all studied models, the QBM/\allowbreak thermometer combination performs similarly well as the exact QBM. The studied models are described in greater detail in Appendix~\ref{sec:systems}. Error bars denote one standard error over five instances.}
        %
        %
        \label{fig:kl_divs}
    \end{center}
\end{figure*}

We notice that, for all considered data distribution instances and models, the QBM/\allowbreak thermometer combination performs similarly to that of an exact QBM with perfect oracle access to quantum thermal states on learning this class of data distributions, even with a finite coherence time. Furthermore, for all considered data distribution instances and models, the QBM/\allowbreak thermometer combination outperforms the classical RBM in KL divergence. For our class of data distributions, the extra connectivity between visible units allowed by QBMs did not offer a significant performance advantage compared to the number of additional trained parameters, as evidenced by empirical measurements of the AIC. Furthermore, the restricted XX model does not seem to perform as well as the restricted transverse Ising model, even though the model is universal for quantum computation~\cite{Cubitt9497}. We believe this is because the upper bound on the loss function given in Eq.~\eqref{eq:u_b_ll} is loose compared to the exact loss function given in Eq.~\eqref{eq:exact_ll}, due to the positive phase of the $XX$ gradient terms vanishing when using Eq.~\eqref{eq:u_b_ll} for training. This could be corrected through the use of relative entropy training~\cite{Kieferova2017,2019arXiv190205162W}, which we will consider in future work.
%
%
%





%
\subsection{Performance Scaling With Noise}\label{sec:noise}

\begin{figure}[ht]
    \begin{center}
        \includegraphics[width=\linewidth]{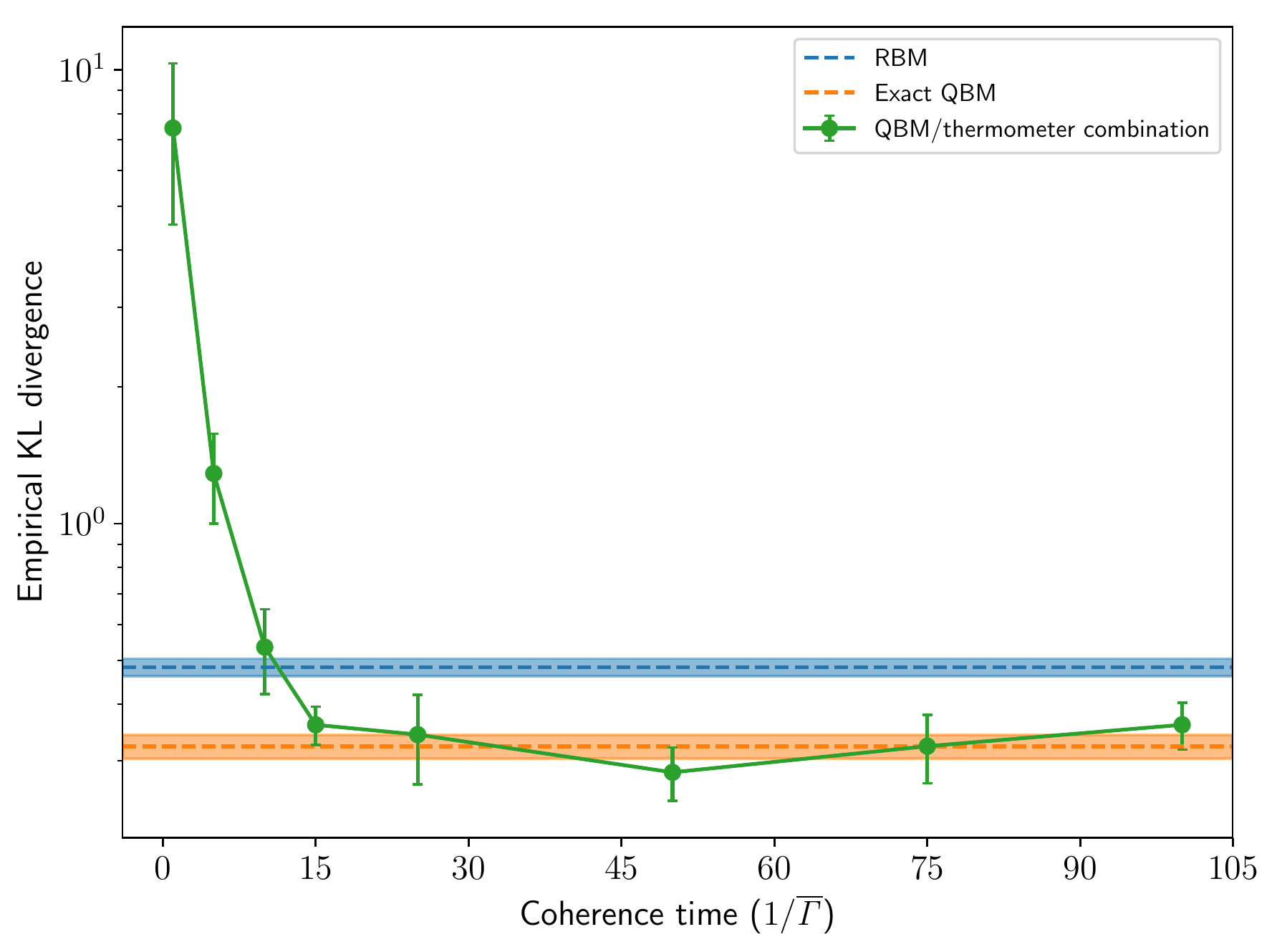}
        \caption{The training performance of our combined QBM/\allowbreak thermometer scheme on six-dimensional Bernoulli mixture models for a restricted transverse Ising model as a function of the system coherence time, in units of the inverse mean single-site transverse field $\frac{1}{\overline{\varGamma}}$ (see Sec.~\ref{sec:noise}). The dashed lines correspond to the mean performance on the same data sets for an RBM and an exact QBM. Above a certain coherence time threshold, the performance of the QBM/\allowbreak thermometer combination is approximately independent of the system noise. The shaded regions and error bars correspond to a confidence interval of one standard error over five instances.}
        %
        %
        %
        %
        \label{fig:kl_div_cts}
    \end{center}
\end{figure}

To test the noise resilience of our training scheme, we tested the performance of our heuristic for multiple simulated coherence times. As shown in Fig.~\ref{fig:kl_div_cts} for the restricted transverse Ising model, the plotted coherence time is $T_1=T_\phi$ in units of $\frac{1}{\overline{\varGamma}}$, where $\overline{\varGamma}$ is the mean single-site transverse field. Above a certain coherence time threshold, the performance of the QBM/\allowbreak thermometer combination is approximately independent of the system noise. For comparison, state of the art neutral atom quantum simulators that naturally implement a similar Hamiltonian achieve both $T_1$ and $T_\phi$ times of approximately $75$ in these units~\cite{Endres1024,PhysRevLett.121.123603}.

\section{Conclusion}\label{sec:conclusion}

In this work we give an efficient heuristic approach for training quantum Boltzmann machines through Hamiltonian simulation based on ETH. As our method relies only on time evolution under ergodic, tunable quantum Hamiltonians, our results can be implemented on NISQ devices~\cite{Preskill2018quantumcomputingin} and quantum simulators. Though there is numerical evidence of the noise-resiliance of our algorithm, further noise mitigation techniques for NISQ devices have been developed and could be deployed for use in conjunction with our algorithm~\cite{PhysRevA.95.042308,PhysRevX.7.021050,PhysRevLett.119.180509,PhysRevA.97.042310,2018arXiv180504492K}. Furthermore, even on small quantum devices, QBMs could work in tandem with classical machine learning architectures to create an architecture that is more expressive than either the small QBM or classical machine learning architecture alone.

The techniques developed in this work may also be useful for the training of general variational quantum algorithms which need to sample many observables in order to evaluate the algorithm's objective function; for example, the Variational Quantum Eigensolver algorithm, which trains on the measured energy of an ansatz state~\cite{peruzzo2014variational}. Instead, one could use temperature measurements of a weakly coupled, small thermometer system as an approximate proxy for these energy evaluations; a course training could begin training on the temperature evaluations, and then refine training with sampling the true objective function. Additionally, one could consider the generalized thermalization of integrable systems to generalized canonical ensembles in training QBMs to take advantage of known symmetries in the data distribution. We hope to explore this and other potential applications in the near future. We also look forward to working with experimental collaborators on potential experimental implementations of this work.

\begin{acknowledgments}
We would like to thank Nathan Wiebe for helpful discussions on this work. We would also like to acknowledge the Zapata Computing scientific team for their insightful and inspiring comments. ERA is partially supported by a Lester Wolfe Fellowship and the Henry W.\ Kendall Fellowship Fund from MIT.
\end{acknowledgments}

\bibliography{main.bib}

\begin{thebibliography}{55}%
\makeatletter
\providecommand \@ifxundefined [1]{%
 \@ifx{#1\undefined}
}%
\providecommand \@ifnum [1]{%
 \ifnum #1\expandafter \@firstoftwo
 \else \expandafter \@secondoftwo
 \fi
}%
\providecommand \@ifx [1]{%
 \ifx #1\expandafter \@firstoftwo
 \else \expandafter \@secondoftwo
 \fi
}%
\providecommand \natexlab [1]{#1}%
\providecommand \enquote  [1]{``#1''}%
\providecommand \bibnamefont  [1]{#1}%
\providecommand \bibfnamefont [1]{#1}%
\providecommand \citenamefont [1]{#1}%
\providecommand \href@noop [0]{\@secondoftwo}%
\providecommand \href [0]{\begingroup \@sanitize@url \@href}%
\providecommand \@href[1]{\@@startlink{#1}\@@href}%
\providecommand \@@href[1]{\endgroup#1\@@endlink}%
\providecommand \@sanitize@url [0]{\catcode `\\12\catcode `\$12\catcode
  `\&12\catcode `\#12\catcode `\^12\catcode `\_12\catcode `\%12\relax}%
\providecommand \@@startlink[1]{}%
\providecommand \@@endlink[0]{}%
\providecommand \url  [0]{\begingroup\@sanitize@url \@url }%
\providecommand \@url [1]{\endgroup\@href {#1}{\urlprefix }}%
\providecommand \urlprefix  [0]{URL }%
\providecommand \Eprint [0]{\href }%
\providecommand \doibase [0]{http://dx.doi.org/}%
\providecommand \selectlanguage [0]{\@gobble}%
\providecommand \bibinfo  [0]{\@secondoftwo}%
\providecommand \bibfield  [0]{\@secondoftwo}%
\providecommand \translation [1]{[#1]}%
\providecommand \BibitemOpen [0]{}%
\providecommand \bibitemStop [0]{}%
\providecommand \bibitemNoStop [0]{.\EOS\space}%
\providecommand \EOS [0]{\spacefactor3000\relax}%
\providecommand \BibitemShut  [1]{\csname bibitem#1\endcsname}%
\let\auto@bib@innerbib\@empty
\bibitem [{\citenamefont {Ackley}\ \emph {et~al.}(1985)\citenamefont {Ackley},
  \citenamefont {Hinton},\ and\ \citenamefont {Sejnowski}}]{ACKLEY1985147}%
  \BibitemOpen
  \bibfield  {author} {\bibinfo {author} {\bibfnamefont {D.~H.}\ \bibnamefont
  {Ackley}}, \bibinfo {author} {\bibfnamefont {G.~E.}\ \bibnamefont {Hinton}},
  \ and\ \bibinfo {author} {\bibfnamefont {T.~J.}\ \bibnamefont {Sejnowski}},\
  }\href {\doibase 10.1016/S0364-0213(85)80012-4} {\bibfield  {journal}
  {\bibinfo  {journal} {Cognitive Sci.}\ }\textbf {\bibinfo {volume} {9}},\
  \bibinfo {pages} {147 } (\bibinfo {year} {1985})}\BibitemShut {NoStop}%
\bibitem [{\citenamefont {Amin}\ \emph {et~al.}(2018)\citenamefont {Amin},
  \citenamefont {Andriyash}, \citenamefont {Rolfe}, \citenamefont
  {Kulchytskyy},\ and\ \citenamefont {Melko}}]{Amin2018}%
  \BibitemOpen
  \bibfield  {author} {\bibinfo {author} {\bibfnamefont {M.~H.}\ \bibnamefont
  {Amin}}, \bibinfo {author} {\bibfnamefont {E.}~\bibnamefont {Andriyash}},
  \bibinfo {author} {\bibfnamefont {J.}~\bibnamefont {Rolfe}}, \bibinfo
  {author} {\bibfnamefont {B.}~\bibnamefont {Kulchytskyy}}, \ and\ \bibinfo
  {author} {\bibfnamefont {R.}~\bibnamefont {Melko}},\ }\href {\doibase
  10.1103/PhysRevX.8.021050} {\bibfield  {journal} {\bibinfo  {journal} {Phys.
  Rev. X}\ }\textbf {\bibinfo {volume} {8}},\ \bibinfo {pages} {021050}
  (\bibinfo {year} {2018})}\BibitemShut {NoStop}%
\bibitem [{\citenamefont {Kieferov{\'{a}}}\ and\ \citenamefont
  {Wiebe}(2017)}]{Kieferova2017}%
  \BibitemOpen
  \bibfield  {author} {\bibinfo {author} {\bibfnamefont {M.}~\bibnamefont
  {Kieferov{\'{a}}}}\ and\ \bibinfo {author} {\bibfnamefont {N.}~\bibnamefont
  {Wiebe}},\ }\href {\doibase 10.1103/PhysRevA.96.062327} {\bibfield  {journal}
  {\bibinfo  {journal} {Phys. Rev. A}\ }\textbf {\bibinfo {volume} {96}},\
  \bibinfo {pages} {062327} (\bibinfo {year} {2017})}\BibitemShut {NoStop}%
\bibitem [{\citenamefont {{Wiebe}}\ \emph {et~al.}(2019)\citenamefont
  {{Wiebe}}, \citenamefont {{Bocharov}}, \citenamefont {{Smolensky}},
  \citenamefont {{Troyer}},\ and\ \citenamefont
  {{Svore}}}]{2019arXiv190205162W}%
  \BibitemOpen
  \bibfield  {author} {\bibinfo {author} {\bibfnamefont {N.}~\bibnamefont
  {{Wiebe}}}, \bibinfo {author} {\bibfnamefont {A.}~\bibnamefont {{Bocharov}}},
  \bibinfo {author} {\bibfnamefont {P.}~\bibnamefont {{Smolensky}}}, \bibinfo
  {author} {\bibfnamefont {M.}~\bibnamefont {{Troyer}}}, \ and\ \bibinfo
  {author} {\bibfnamefont {K.~M.}\ \bibnamefont {{Svore}}},\ }\href@noop {}
  {\enquote {\bibinfo {title} {{Quantum Language Processing}},}\ } (\bibinfo
  {year} {2019}),\ \Eprint {http://arxiv.org/abs/1902.05162} {arXiv:1902.05162
  [quant-ph]} \BibitemShut {NoStop}%
\bibitem [{\citenamefont {Amin}(2015)}]{PhysRevA.92.052323}%
  \BibitemOpen
  \bibfield  {author} {\bibinfo {author} {\bibfnamefont {M.~H.}\ \bibnamefont
  {Amin}},\ }\href {\doibase 10.1103/PhysRevA.92.052323} {\bibfield  {journal}
  {\bibinfo  {journal} {Phys. Rev. A}\ }\textbf {\bibinfo {volume} {92}},\
  \bibinfo {pages} {052323} (\bibinfo {year} {2015})}\BibitemShut {NoStop}%
\bibitem [{\citenamefont {Mandr\`a}\ \emph {et~al.}(2016)\citenamefont
  {Mandr\`a}, \citenamefont {Zhu}, \citenamefont {Wang}, \citenamefont
  {Perdomo-Ortiz},\ and\ \citenamefont
  {Katzgraber}}]{Mandra2016StrengthsApproaches}%
  \BibitemOpen
  \bibfield  {author} {\bibinfo {author} {\bibfnamefont {S.}~\bibnamefont
  {Mandr\`a}}, \bibinfo {author} {\bibfnamefont {Z.}~\bibnamefont {Zhu}},
  \bibinfo {author} {\bibfnamefont {W.}~\bibnamefont {Wang}}, \bibinfo {author}
  {\bibfnamefont {A.}~\bibnamefont {Perdomo-Ortiz}}, \ and\ \bibinfo {author}
  {\bibfnamefont {H.~G.}\ \bibnamefont {Katzgraber}},\ }\href {\doibase
  10.1103/PhysRevA.94.022337} {\bibfield  {journal} {\bibinfo  {journal} {Phys.
  Rev. A}\ }\textbf {\bibinfo {volume} {94}},\ \bibinfo {pages} {022337}
  (\bibinfo {year} {2016})}\BibitemShut {NoStop}%
\bibitem [{\citenamefont {{Verdon}}\ \emph {et~al.}(2017)\citenamefont
  {{Verdon}}, \citenamefont {{Broughton}},\ and\ \citenamefont
  {{Biamonte}}}]{Verdon2017ACircuits}%
  \BibitemOpen
  \bibfield  {author} {\bibinfo {author} {\bibfnamefont {G.}~\bibnamefont
  {{Verdon}}}, \bibinfo {author} {\bibfnamefont {M.}~\bibnamefont
  {{Broughton}}}, \ and\ \bibinfo {author} {\bibfnamefont {J.}~\bibnamefont
  {{Biamonte}}},\ }\href@noop {} {\enquote {\bibinfo {title} {{A quantum
  algorithm to train neural networks using low-depth circuits}},}\ } (\bibinfo
  {year} {2017}),\ \Eprint {http://arxiv.org/abs/1712.05304} {arXiv:1712.05304
  [quant-ph]} \BibitemShut {NoStop}%
\bibitem [{\citenamefont {{Wu}}\ and\ \citenamefont
  {{Hsieh}}(2018)}]{2018arXiv181111756W}%
  \BibitemOpen
  \bibfield  {author} {\bibinfo {author} {\bibfnamefont {J.}~\bibnamefont
  {{Wu}}}\ and\ \bibinfo {author} {\bibfnamefont {T.~H.}\ \bibnamefont
  {{Hsieh}}},\ }\href@noop {} {\enquote {\bibinfo {title} {{Variational Thermal
  Quantum Simulation via Thermofield Double States}},}\ } (\bibinfo {year}
  {2018}),\ \Eprint {http://arxiv.org/abs/1811.11756} {arXiv:1811.11756
  [cond-mat.str-el]} \BibitemShut {NoStop}%
\bibitem [{\citenamefont {Farhi}\ \emph {et~al.}(2014)\citenamefont {Farhi},
  \citenamefont {Goldstone},\ and\ \citenamefont {Gutmann}}]{Farhi2014}%
  \BibitemOpen
  \bibfield  {author} {\bibinfo {author} {\bibfnamefont {E.}~\bibnamefont
  {Farhi}}, \bibinfo {author} {\bibfnamefont {J.}~\bibnamefont {Goldstone}}, \
  and\ \bibinfo {author} {\bibfnamefont {S.}~\bibnamefont {Gutmann}},\
  }\href@noop {} {\enquote {\bibinfo {title} {{A Quantum Approximate
  Optimization Algorithm}},}\ } (\bibinfo {year} {2014}),\ \Eprint
  {http://arxiv.org/abs/1411.4028} {arXiv:1411.4028 [quant-ph]} \BibitemShut
  {NoStop}%
\bibitem [{\citenamefont {{McArdle}}\ \emph {et~al.}(2018)\citenamefont
  {{McArdle}}, \citenamefont {{Jones}}, \citenamefont {{Endo}}, \citenamefont
  {{Li}}, \citenamefont {{Benjamin}},\ and\ \citenamefont
  {{Yuan}}}]{2018arXiv180403023M}%
  \BibitemOpen
  \bibfield  {author} {\bibinfo {author} {\bibfnamefont {S.}~\bibnamefont
  {{McArdle}}}, \bibinfo {author} {\bibfnamefont {T.}~\bibnamefont {{Jones}}},
  \bibinfo {author} {\bibfnamefont {S.}~\bibnamefont {{Endo}}}, \bibinfo
  {author} {\bibfnamefont {Y.}~\bibnamefont {{Li}}}, \bibinfo {author}
  {\bibfnamefont {S.}~\bibnamefont {{Benjamin}}}, \ and\ \bibinfo {author}
  {\bibfnamefont {X.}~\bibnamefont {{Yuan}}},\ }\href@noop {} {\enquote
  {\bibinfo {title} {{Variational quantum simulation of imaginary time
  evolution}},}\ } (\bibinfo {year} {2018}),\ \Eprint
  {http://arxiv.org/abs/1804.03023} {arXiv:1804.03023 [quant-ph]} \BibitemShut
  {NoStop}%
\bibitem [{\citenamefont {{Martyn}}\ and\ \citenamefont
  {{Swingle}}(2018)}]{2018arXiv181201015M}%
  \BibitemOpen
  \bibfield  {author} {\bibinfo {author} {\bibfnamefont {J.}~\bibnamefont
  {{Martyn}}}\ and\ \bibinfo {author} {\bibfnamefont {B.}~\bibnamefont
  {{Swingle}}},\ }\href@noop {} {\enquote {\bibinfo {title} {{Product Spectrum
  Ansatz and the Simplicity of Thermal States}},}\ } (\bibinfo {year} {2018}),\
  \Eprint {http://arxiv.org/abs/1812.01015} {arXiv:1812.01015
  [cond-mat.str-el]} \BibitemShut {NoStop}%
\bibitem [{\citenamefont {Poulin}\ and\ \citenamefont
  {Wocjan}(2009)}]{PhysRevLett.103.220502}%
  \BibitemOpen
  \bibfield  {author} {\bibinfo {author} {\bibfnamefont {D.}~\bibnamefont
  {Poulin}}\ and\ \bibinfo {author} {\bibfnamefont {P.}~\bibnamefont
  {Wocjan}},\ }\href {\doibase 10.1103/PhysRevLett.103.220502} {\bibfield
  {journal} {\bibinfo  {journal} {Phys. Rev. Lett.}\ }\textbf {\bibinfo
  {volume} {103}},\ \bibinfo {pages} {220502} (\bibinfo {year}
  {2009})}\BibitemShut {NoStop}%
\bibitem [{\citenamefont {Yung}\ and\ \citenamefont
  {Aspuru-Guzik}(2012)}]{Yung754}%
  \BibitemOpen
  \bibfield  {author} {\bibinfo {author} {\bibfnamefont {M.-H.}\ \bibnamefont
  {Yung}}\ and\ \bibinfo {author} {\bibfnamefont {A.}~\bibnamefont
  {Aspuru-Guzik}},\ }\href {\doibase 10.1073/pnas.1111758109} {\bibfield
  {journal} {\bibinfo  {journal} {Proc. Natl. Acad. Sci. USA}\ }\textbf
  {\bibinfo {volume} {109}},\ \bibinfo {pages} {754} (\bibinfo {year}
  {2012})}\BibitemShut {NoStop}%
\bibitem [{\citenamefont {{Narayan Chowdhury}}\ and\ \citenamefont
  {{Somma}}(2016)}]{2016arXiv160302940N}%
  \BibitemOpen
  \bibfield  {author} {\bibinfo {author} {\bibfnamefont {A.}~\bibnamefont
  {{Narayan Chowdhury}}}\ and\ \bibinfo {author} {\bibfnamefont {R.~D.}\
  \bibnamefont {{Somma}}},\ }\href@noop {} {\enquote {\bibinfo {title}
  {{Quantum algorithms for Gibbs sampling and hitting-time estimation}},}\ }
  (\bibinfo {year} {2016}),\ \Eprint {http://arxiv.org/abs/1603.02940}
  {arXiv:1603.02940 [quant-ph]} \BibitemShut {NoStop}%
\bibitem [{\citenamefont {{Wiebe}}\ \emph {et~al.}(2015)\citenamefont
  {{Wiebe}}, \citenamefont {{Kapoor}}, \citenamefont {{Granade}},\ and\
  \citenamefont {{Svore}}}]{2015arXiv150702642W}%
  \BibitemOpen
  \bibfield  {author} {\bibinfo {author} {\bibfnamefont {N.}~\bibnamefont
  {{Wiebe}}}, \bibinfo {author} {\bibfnamefont {A.}~\bibnamefont {{Kapoor}}},
  \bibinfo {author} {\bibfnamefont {C.}~\bibnamefont {{Granade}}}, \ and\
  \bibinfo {author} {\bibfnamefont {K.~M.}\ \bibnamefont {{Svore}}},\
  }\href@noop {} {\enquote {\bibinfo {title} {{Quantum Inspired Training for
  Boltzmann Machines}},}\ } (\bibinfo {year} {2015}),\ \Eprint
  {http://arxiv.org/abs/1507.02642} {arXiv:1507.02642 [cs.LG]} \BibitemShut
  {NoStop}%
\bibitem [{\citenamefont {{Motta}}\ \emph {et~al.}(2019)\citenamefont
  {{Motta}}, \citenamefont {{Sun}}, \citenamefont {{Teck Keng Tan}},
  \citenamefont {{O' Rourke}}, \citenamefont {{Ye}}, \citenamefont {{Minnich}},
  \citenamefont {{Brandao}},\ and\ \citenamefont {{Kin-Lic
  Chan}}}]{2019arXiv190107653M}%
  \BibitemOpen
  \bibfield  {author} {\bibinfo {author} {\bibfnamefont {M.}~\bibnamefont
  {{Motta}}}, \bibinfo {author} {\bibfnamefont {C.}~\bibnamefont {{Sun}}},
  \bibinfo {author} {\bibfnamefont {A.}~\bibnamefont {{Teck Keng Tan}}},
  \bibinfo {author} {\bibfnamefont {M.~J.}\ \bibnamefont {{O' Rourke}}},
  \bibinfo {author} {\bibfnamefont {E.}~\bibnamefont {{Ye}}}, \bibinfo {author}
  {\bibfnamefont {A.~J.}\ \bibnamefont {{Minnich}}}, \bibinfo {author}
  {\bibfnamefont {F.~G.~S.~L.}\ \bibnamefont {{Brandao}}}, \ and\ \bibinfo
  {author} {\bibfnamefont {G.}~\bibnamefont {{Kin-Lic Chan}}},\ }\href@noop {}
  {\enquote {\bibinfo {title} {{Quantum Imaginary Time Evolution, Quantum
  Lanczos, and Quantum Thermal Averaging}},}\ } (\bibinfo {year} {2019}),\
  \Eprint {http://arxiv.org/abs/1901.07653} {arXiv:1901.07653 [quant-ph]}
  \BibitemShut {NoStop}%
\bibitem [{\citenamefont {Srednicki}(1994)}]{PhysRevE.50.888}%
  \BibitemOpen
  \bibfield  {author} {\bibinfo {author} {\bibfnamefont {M.}~\bibnamefont
  {Srednicki}},\ }\href {\doibase 10.1103/PhysRevE.50.888} {\bibfield
  {journal} {\bibinfo  {journal} {Phys. Rev. E}\ }\textbf {\bibinfo {volume}
  {50}},\ \bibinfo {pages} {888} (\bibinfo {year} {1994})}\BibitemShut
  {NoStop}%
\bibitem [{\citenamefont {Deutsch}(1991)}]{PhysRevA.43.2046}%
  \BibitemOpen
  \bibfield  {author} {\bibinfo {author} {\bibfnamefont {J.~M.}\ \bibnamefont
  {Deutsch}},\ }\href {\doibase 10.1103/PhysRevA.43.2046} {\bibfield  {journal}
  {\bibinfo  {journal} {Phys. Rev. A}\ }\textbf {\bibinfo {volume} {43}},\
  \bibinfo {pages} {2046} (\bibinfo {year} {1991})}\BibitemShut {NoStop}%
\bibitem [{\citenamefont {D'Alessio}\ \emph {et~al.}(2016)\citenamefont
  {D'Alessio}, \citenamefont {Kafri}, \citenamefont {Polkovnikov},\ and\
  \citenamefont {Rigol}}]{DAlessio2016}%
  \BibitemOpen
  \bibfield  {author} {\bibinfo {author} {\bibfnamefont {L.}~\bibnamefont
  {D'Alessio}}, \bibinfo {author} {\bibfnamefont {Y.}~\bibnamefont {Kafri}},
  \bibinfo {author} {\bibfnamefont {A.}~\bibnamefont {Polkovnikov}}, \ and\
  \bibinfo {author} {\bibfnamefont {M.}~\bibnamefont {Rigol}},\ }\href
  {\doibase 10.1080/00018732.2016.1198134} {\bibfield  {journal} {\bibinfo
  {journal} {Adv. Phys.}\ }\textbf {\bibinfo {volume} {65}},\ \bibinfo {pages}
  {239} (\bibinfo {year} {2016})}\BibitemShut {NoStop}%
\bibitem [{\citenamefont {Rigol}\ \emph {et~al.}(2008)\citenamefont {Rigol},
  \citenamefont {Dunjko},\ and\ \citenamefont
  {Olshanii}}]{rigol2008thermalization}%
  \BibitemOpen
  \bibfield  {author} {\bibinfo {author} {\bibfnamefont {M.}~\bibnamefont
  {Rigol}}, \bibinfo {author} {\bibfnamefont {V.}~\bibnamefont {Dunjko}}, \
  and\ \bibinfo {author} {\bibfnamefont {M.}~\bibnamefont {Olshanii}},\ }\href
  {\doibase 10.1038/nature06838} {\bibfield  {journal} {\bibinfo  {journal}
  {Nature}\ }\textbf {\bibinfo {volume} {452}},\ \bibinfo {pages} {854}
  (\bibinfo {year} {2008})}\BibitemShut {NoStop}%
\bibitem [{\citenamefont {Mondaini}\ \emph {et~al.}(2016)\citenamefont
  {Mondaini}, \citenamefont {Fratus}, \citenamefont {Srednicki},\ and\
  \citenamefont {Rigol}}]{PhysRevE.93.032104}%
  \BibitemOpen
  \bibfield  {author} {\bibinfo {author} {\bibfnamefont {R.}~\bibnamefont
  {Mondaini}}, \bibinfo {author} {\bibfnamefont {K.~R.}\ \bibnamefont
  {Fratus}}, \bibinfo {author} {\bibfnamefont {M.}~\bibnamefont {Srednicki}}, \
  and\ \bibinfo {author} {\bibfnamefont {M.}~\bibnamefont {Rigol}},\ }\href
  {\doibase 10.1103/PhysRevE.93.032104} {\bibfield  {journal} {\bibinfo
  {journal} {Phys. Rev. E}\ }\textbf {\bibinfo {volume} {93}},\ \bibinfo
  {pages} {032104} (\bibinfo {year} {2016})}\BibitemShut {NoStop}%
\bibitem [{\citenamefont {Garrison}\ and\ \citenamefont
  {Grover}(2018)}]{Garrison2018}%
  \BibitemOpen
  \bibfield  {author} {\bibinfo {author} {\bibfnamefont {J.~R.}\ \bibnamefont
  {Garrison}}\ and\ \bibinfo {author} {\bibfnamefont {T.}~\bibnamefont
  {Grover}},\ }\href {\doibase 10.1103/PhysRevX.8.021026} {\bibfield  {journal}
  {\bibinfo  {journal} {Phys. Rev. X}\ }\textbf {\bibinfo {volume} {8}},\
  \bibinfo {pages} {021026} (\bibinfo {year} {2018})}\BibitemShut {NoStop}%
\bibitem [{\citenamefont {Neill}\ \emph {et~al.}(2016)\citenamefont {Neill},
  \citenamefont {Roushan}, \citenamefont {Fang}, \citenamefont {Chen},
  \citenamefont {Kolodrubetz}, \citenamefont {Chen}, \citenamefont {Megrant},
  \citenamefont {Barends}, \citenamefont {Campbell}, \citenamefont {Chiaro}
  \emph {et~al.}}]{neill2016ergodic}%
  \BibitemOpen
  \bibfield  {author} {\bibinfo {author} {\bibfnamefont {C.}~\bibnamefont
  {Neill}}, \bibinfo {author} {\bibfnamefont {P.}~\bibnamefont {Roushan}},
  \bibinfo {author} {\bibfnamefont {M.}~\bibnamefont {Fang}}, \bibinfo {author}
  {\bibfnamefont {Y.}~\bibnamefont {Chen}}, \bibinfo {author} {\bibfnamefont
  {M.}~\bibnamefont {Kolodrubetz}}, \bibinfo {author} {\bibfnamefont
  {Z.}~\bibnamefont {Chen}}, \bibinfo {author} {\bibfnamefont {A.}~\bibnamefont
  {Megrant}}, \bibinfo {author} {\bibfnamefont {R.}~\bibnamefont {Barends}},
  \bibinfo {author} {\bibfnamefont {B.}~\bibnamefont {Campbell}}, \bibinfo
  {author} {\bibfnamefont {B.}~\bibnamefont {Chiaro}},  \emph {et~al.},\ }\href
  {\doibase 10.1038/nphys3830} {\bibfield  {journal} {\bibinfo  {journal} {Nat.
  Phys.}\ }\textbf {\bibinfo {volume} {12}},\ \bibinfo {pages} {1037} (\bibinfo
  {year} {2016})}\BibitemShut {NoStop}%
\bibitem [{\citenamefont {Kaufman}\ \emph {et~al.}(2016)\citenamefont
  {Kaufman}, \citenamefont {Tai}, \citenamefont {Lukin}, \citenamefont
  {Rispoli}, \citenamefont {Schittko}, \citenamefont {Preiss},\ and\
  \citenamefont {Greiner}}]{Kaufman794}%
  \BibitemOpen
  \bibfield  {author} {\bibinfo {author} {\bibfnamefont {A.~M.}\ \bibnamefont
  {Kaufman}}, \bibinfo {author} {\bibfnamefont {M.~E.}\ \bibnamefont {Tai}},
  \bibinfo {author} {\bibfnamefont {A.}~\bibnamefont {Lukin}}, \bibinfo
  {author} {\bibfnamefont {M.}~\bibnamefont {Rispoli}}, \bibinfo {author}
  {\bibfnamefont {R.}~\bibnamefont {Schittko}}, \bibinfo {author}
  {\bibfnamefont {P.~M.}\ \bibnamefont {Preiss}}, \ and\ \bibinfo {author}
  {\bibfnamefont {M.}~\bibnamefont {Greiner}},\ }\href {\doibase
  10.1126/science.aaf6725} {\bibfield  {journal} {\bibinfo  {journal}
  {Science}\ }\textbf {\bibinfo {volume} {353}},\ \bibinfo {pages} {794}
  (\bibinfo {year} {2016})}\BibitemShut {NoStop}%
\bibitem [{\citenamefont {Hinton}(2002)}]{doi:10.1162/089976602760128018}%
  \BibitemOpen
  \bibfield  {author} {\bibinfo {author} {\bibfnamefont {G.~E.}\ \bibnamefont
  {Hinton}},\ }\href {\doibase 10.1162/089976602760128018} {\bibfield
  {journal} {\bibinfo  {journal} {Neural Comput.}\ }\textbf {\bibinfo {volume}
  {14}},\ \bibinfo {pages} {1771} (\bibinfo {year} {2002})}\BibitemShut
  {NoStop}%
\bibitem [{\citenamefont {Biamonte}\ and\ \citenamefont
  {Love}(2008)}]{PhysRevA.78.012352}%
  \BibitemOpen
  \bibfield  {author} {\bibinfo {author} {\bibfnamefont {J.~D.}\ \bibnamefont
  {Biamonte}}\ and\ \bibinfo {author} {\bibfnamefont {P.~J.}\ \bibnamefont
  {Love}},\ }\href {\doibase 10.1103/PhysRevA.78.012352} {\bibfield  {journal}
  {\bibinfo  {journal} {Phys. Rev. A}\ }\textbf {\bibinfo {volume} {78}},\
  \bibinfo {pages} {012352} (\bibinfo {year} {2008})}\BibitemShut {NoStop}%
\bibitem [{\citenamefont {{Barahona}}(1982)}]{1982JPhA...15.3241B}%
  \BibitemOpen
  \bibfield  {author} {\bibinfo {author} {\bibfnamefont {F.}~\bibnamefont
  {{Barahona}}},\ }\href {\doibase 10.1088/0305-4470/15/10/028} {\bibfield
  {journal} {\bibinfo  {journal} {J. Phys. A}\ }\textbf {\bibinfo {volume}
  {15}},\ \bibinfo {pages} {3241} (\bibinfo {year} {1982})}\BibitemShut
  {NoStop}%
\bibitem [{\citenamefont {Srednicki}(1999)}]{Srednicki_1999}%
  \BibitemOpen
  \bibfield  {author} {\bibinfo {author} {\bibfnamefont {M.}~\bibnamefont
  {Srednicki}},\ }\href {\doibase 10.1088/0305-4470/32/7/007} {\bibfield
  {journal} {\bibinfo  {journal} {J. Phys. A}\ }\textbf {\bibinfo {volume}
  {32}},\ \bibinfo {pages} {1163} (\bibinfo {year} {1999})}\BibitemShut
  {NoStop}%
\bibitem [{\citenamefont {Touchette}\ \emph {et~al.}(2004)\citenamefont
  {Touchette}, \citenamefont {Ellis},\ and\ \citenamefont
  {Turkington}}]{TOUCHETTE2004138}%
  \BibitemOpen
  \bibfield  {author} {\bibinfo {author} {\bibfnamefont {H.}~\bibnamefont
  {Touchette}}, \bibinfo {author} {\bibfnamefont {R.~S.}\ \bibnamefont
  {Ellis}}, \ and\ \bibinfo {author} {\bibfnamefont {B.}~\bibnamefont
  {Turkington}},\ }\href {\doibase 10.1016/j.physa.2004.03.088} {\bibfield
  {journal} {\bibinfo  {journal} {Physica A}\ }\textbf {\bibinfo {volume}
  {340}},\ \bibinfo {pages} {138 } (\bibinfo {year} {2004})}\BibitemShut
  {NoStop}%
\bibitem [{\citenamefont {Ellis}\ \emph {et~al.}(2000)\citenamefont {Ellis},
  \citenamefont {Haven},\ and\ \citenamefont {Turkington}}]{Ellis2000}%
  \BibitemOpen
  \bibfield  {author} {\bibinfo {author} {\bibfnamefont {R.~S.}\ \bibnamefont
  {Ellis}}, \bibinfo {author} {\bibfnamefont {K.}~\bibnamefont {Haven}}, \ and\
  \bibinfo {author} {\bibfnamefont {B.}~\bibnamefont {Turkington}},\ }\href
  {\doibase 10.1023/A:1026446225804} {\bibfield  {journal} {\bibinfo  {journal}
  {J. Stat. Phys.}\ }\textbf {\bibinfo {volume} {101}},\ \bibinfo {pages} {999}
  (\bibinfo {year} {2000})}\BibitemShut {NoStop}%
\bibitem [{Note1()}]{Note1}%
  \BibitemOpen
  \bibinfo {note} {Though this is guaranteed only asymptotically, numerically
  we find this to be true even for relatively small system sizes (see Sec.~\ref
  {sec:ns}).}\BibitemShut {Stop}%
\bibitem [{\citenamefont {Dymarsky}\ \emph {et~al.}(2018)\citenamefont
  {Dymarsky}, \citenamefont {Lashkari},\ and\ \citenamefont
  {Liu}}]{PhysRevE.97.012140}%
  \BibitemOpen
  \bibfield  {author} {\bibinfo {author} {\bibfnamefont {A.}~\bibnamefont
  {Dymarsky}}, \bibinfo {author} {\bibfnamefont {N.}~\bibnamefont {Lashkari}},
  \ and\ \bibinfo {author} {\bibfnamefont {H.}~\bibnamefont {Liu}},\ }\href
  {\doibase 10.1103/PhysRevE.97.012140} {\bibfield  {journal} {\bibinfo
  {journal} {Phys. Rev. E}\ }\textbf {\bibinfo {volume} {97}},\ \bibinfo
  {pages} {012140} (\bibinfo {year} {2018})}\BibitemShut {NoStop}%
\bibitem [{\citenamefont {Costeniuc}\ \emph {et~al.}(2005)\citenamefont
  {Costeniuc}, \citenamefont {Ellis}, \citenamefont {Touchette},\ and\
  \citenamefont {Turkington}}]{Costeniuc2005}%
  \BibitemOpen
  \bibfield  {author} {\bibinfo {author} {\bibfnamefont {M.}~\bibnamefont
  {Costeniuc}}, \bibinfo {author} {\bibfnamefont {R.~S.}\ \bibnamefont
  {Ellis}}, \bibinfo {author} {\bibfnamefont {H.}~\bibnamefont {Touchette}}, \
  and\ \bibinfo {author} {\bibfnamefont {B.}~\bibnamefont {Turkington}},\
  }\href {\doibase 10.1007/s10955-005-4407-0} {\bibfield  {journal} {\bibinfo
  {journal} {J. Stat. Phys.}\ }\textbf {\bibinfo {volume} {119}},\ \bibinfo
  {pages} {1283} (\bibinfo {year} {2005})}\BibitemShut {NoStop}%
\bibitem [{\citenamefont {Serbyn}\ \emph {et~al.}(2013)\citenamefont {Serbyn},
  \citenamefont {Papi\ifmmode~\acute{c}\else \'{c}\fi{}},\ and\ \citenamefont
  {Abanin}}]{PhysRevLett.111.127201}%
  \BibitemOpen
  \bibfield  {author} {\bibinfo {author} {\bibfnamefont {M.}~\bibnamefont
  {Serbyn}}, \bibinfo {author} {\bibfnamefont {Z.}~\bibnamefont
  {Papi\ifmmode~\acute{c}\else \'{c}\fi{}}}, \ and\ \bibinfo {author}
  {\bibfnamefont {D.~A.}\ \bibnamefont {Abanin}},\ }\href {\doibase
  10.1103/PhysRevLett.111.127201} {\bibfield  {journal} {\bibinfo  {journal}
  {Phys. Rev. Lett.}\ }\textbf {\bibinfo {volume} {111}},\ \bibinfo {pages}
  {127201} (\bibinfo {year} {2013})}\BibitemShut {NoStop}%
\bibitem [{Note2()}]{Note2}%
  \BibitemOpen
  \bibinfo {note} {Note, however, that the integrabiltiy restriction described
  in Sec.~\ref {sec:real_eth} means that this procedure has no equivalent in
  the classical regime. For a diagonal Hamiltonian, all observables that must
  be sampled in training commute with the Hamiltonian and are therefore
  conserved; therefore, ETH only implies thermalization up to to the local
  conservation of spin, and sampling diagonal observables in the time-evolved
  state will then be equivalent to sampling those observables in the initial
  state. See our brief discussion in Sec.~\ref {sec:real_eth} on generalized
  canonical ensembles for details.}\BibitemShut {Stop}%
\bibitem [{Note3()}]{Note3}%
  \BibitemOpen
  \bibinfo {note} {This is assuming all components of the $k$th moment are
  independent; in general, if there are $\iota $ independent components of the
  $k$th moment, we must have that $m=O\left (n^{\iota \left (k\right )}\right
  )$.}\BibitemShut {Stop}%
\bibitem [{\citenamefont {Tchebycheff}(1890)}]{tchebycheff1890}%
  \BibitemOpen
  \bibfield  {author} {\bibinfo {author} {\bibfnamefont {P.}~\bibnamefont
  {Tchebycheff}},\ }\href {\doibase 10.1007/BF02413327} {\bibfield  {journal}
  {\bibinfo  {journal} {Acta Math.}\ }\textbf {\bibinfo {volume} {14}},\
  \bibinfo {pages} {305} (\bibinfo {year} {1890})}\BibitemShut {NoStop}%
\bibitem [{\citenamefont {{Juan}}\ and\ \citenamefont
  {{Vidal}}(2004)}]{1334543}%
  \BibitemOpen
  \bibfield  {author} {\bibinfo {author} {\bibfnamefont {A.}~\bibnamefont
  {{Juan}}}\ and\ \bibinfo {author} {\bibfnamefont {E.}~\bibnamefont
  {{Vidal}}},\ }in\ \href {\doibase 10.1109/ICPR.2004.1334543} {\emph {\bibinfo
  {booktitle} {Proceedings of the 17th International Conference on Pattern
  Recognition, 2004. ICPR 2004.}}},\ Vol.~\bibinfo {volume} {3}\ (\bibinfo
  {year} {2004})\ pp.\ \bibinfo {pages} {367--370}\BibitemShut {NoStop}%
\bibitem [{\citenamefont {Berry}\ and\ \citenamefont
  {Robnik}(1984)}]{Berry_1984}%
  \BibitemOpen
  \bibfield  {author} {\bibinfo {author} {\bibfnamefont {M.~V.}\ \bibnamefont
  {Berry}}\ and\ \bibinfo {author} {\bibfnamefont {M.}~\bibnamefont {Robnik}},\
  }\href {\doibase 10.1088/0305-4470/17/12/013} {\bibfield  {journal} {\bibinfo
   {journal} {J. Phys. A}\ }\textbf {\bibinfo {volume} {17}},\ \bibinfo {pages}
  {2413} (\bibinfo {year} {1984})}\BibitemShut {NoStop}%
\bibitem [{\citenamefont {Kullback}\ and\ \citenamefont
  {Leibler}(1951)}]{kullback1951}%
  \BibitemOpen
  \bibfield  {author} {\bibinfo {author} {\bibfnamefont {S.}~\bibnamefont
  {Kullback}}\ and\ \bibinfo {author} {\bibfnamefont {R.~A.}\ \bibnamefont
  {Leibler}},\ }\href {\doibase 10.1214/aoms/1177729694} {\bibfield  {journal}
  {\bibinfo  {journal} {Ann. Math. Statist.}\ }\textbf {\bibinfo {volume}
  {22}},\ \bibinfo {pages} {79} (\bibinfo {year} {1951})}\BibitemShut {NoStop}%
\bibitem [{\citenamefont {Akaike}(1974)}]{Akaike1974}%
  \BibitemOpen
  \bibfield  {author} {\bibinfo {author} {\bibfnamefont {H.}~\bibnamefont
  {Akaike}},\ }\href {\doibase 10.1109/tac.1974.1100705} {\bibfield  {journal}
  {\bibinfo  {journal} {IEEE T. Automat. Contr.}\ }\textbf {\bibinfo {volume}
  {19}},\ \bibinfo {pages} {716} (\bibinfo {year} {1974})}\BibitemShut
  {NoStop}%
\bibitem [{\citenamefont {Cubitt}\ \emph {et~al.}(2018)\citenamefont {Cubitt},
  \citenamefont {Montanaro},\ and\ \citenamefont {Piddock}}]{Cubitt9497}%
  \BibitemOpen
  \bibfield  {author} {\bibinfo {author} {\bibfnamefont {T.~S.}\ \bibnamefont
  {Cubitt}}, \bibinfo {author} {\bibfnamefont {A.}~\bibnamefont {Montanaro}}, \
  and\ \bibinfo {author} {\bibfnamefont {S.}~\bibnamefont {Piddock}},\ }\href
  {\doibase 10.1073/pnas.1804949115} {\bibfield  {journal} {\bibinfo  {journal}
  {Proc. Natl. Acad. Sci. USA}\ }\textbf {\bibinfo {volume} {115}},\ \bibinfo
  {pages} {9497} (\bibinfo {year} {2018})}\BibitemShut {NoStop}%
\bibitem [{\citenamefont {Endres}\ \emph {et~al.}(2016)\citenamefont {Endres},
  \citenamefont {Bernien}, \citenamefont {Keesling}, \citenamefont {Levine},
  \citenamefont {Anschuetz}, \citenamefont {Krajenbrink}, \citenamefont
  {Senko}, \citenamefont {Vuletic}, \citenamefont {Greiner},\ and\
  \citenamefont {Lukin}}]{Endres1024}%
  \BibitemOpen
  \bibfield  {author} {\bibinfo {author} {\bibfnamefont {M.}~\bibnamefont
  {Endres}}, \bibinfo {author} {\bibfnamefont {H.}~\bibnamefont {Bernien}},
  \bibinfo {author} {\bibfnamefont {A.}~\bibnamefont {Keesling}}, \bibinfo
  {author} {\bibfnamefont {H.}~\bibnamefont {Levine}}, \bibinfo {author}
  {\bibfnamefont {E.~R.}\ \bibnamefont {Anschuetz}}, \bibinfo {author}
  {\bibfnamefont {A.}~\bibnamefont {Krajenbrink}}, \bibinfo {author}
  {\bibfnamefont {C.}~\bibnamefont {Senko}}, \bibinfo {author} {\bibfnamefont
  {V.}~\bibnamefont {Vuletic}}, \bibinfo {author} {\bibfnamefont
  {M.}~\bibnamefont {Greiner}}, \ and\ \bibinfo {author} {\bibfnamefont
  {M.~D.}\ \bibnamefont {Lukin}},\ }\href {\doibase 10.1126/science.aah3752}
  {\bibfield  {journal} {\bibinfo  {journal} {Science}\ }\textbf {\bibinfo
  {volume} {354}},\ \bibinfo {pages} {1024} (\bibinfo {year}
  {2016})}\BibitemShut {NoStop}%
\bibitem [{\citenamefont {Levine}\ \emph {et~al.}(2018)\citenamefont {Levine},
  \citenamefont {Keesling}, \citenamefont {Omran}, \citenamefont {Bernien},
  \citenamefont {Schwartz}, \citenamefont {Zibrov}, \citenamefont {Endres},
  \citenamefont {Greiner}, \citenamefont {Vuleti\ifmmode~\acute{c}\else
  \'{c}\fi{}},\ and\ \citenamefont {Lukin}}]{PhysRevLett.121.123603}%
  \BibitemOpen
  \bibfield  {author} {\bibinfo {author} {\bibfnamefont {H.}~\bibnamefont
  {Levine}}, \bibinfo {author} {\bibfnamefont {A.}~\bibnamefont {Keesling}},
  \bibinfo {author} {\bibfnamefont {A.}~\bibnamefont {Omran}}, \bibinfo
  {author} {\bibfnamefont {H.}~\bibnamefont {Bernien}}, \bibinfo {author}
  {\bibfnamefont {S.}~\bibnamefont {Schwartz}}, \bibinfo {author}
  {\bibfnamefont {A.~S.}\ \bibnamefont {Zibrov}}, \bibinfo {author}
  {\bibfnamefont {M.}~\bibnamefont {Endres}}, \bibinfo {author} {\bibfnamefont
  {M.}~\bibnamefont {Greiner}}, \bibinfo {author} {\bibfnamefont
  {V.}~\bibnamefont {Vuleti\ifmmode~\acute{c}\else \'{c}\fi{}}}, \ and\
  \bibinfo {author} {\bibfnamefont {M.~D.}\ \bibnamefont {Lukin}},\ }\href
  {\doibase 10.1103/PhysRevLett.121.123603} {\bibfield  {journal} {\bibinfo
  {journal} {Phys. Rev. Lett.}\ }\textbf {\bibinfo {volume} {121}},\ \bibinfo
  {pages} {123603} (\bibinfo {year} {2018})}\BibitemShut {NoStop}%
\bibitem [{\citenamefont {Preskill}(2018)}]{Preskill2018quantumcomputingin}%
  \BibitemOpen
  \bibfield  {author} {\bibinfo {author} {\bibfnamefont {J.}~\bibnamefont
  {Preskill}},\ }\href {\doibase 10.22331/q-2018-08-06-79} {\bibfield
  {journal} {\bibinfo  {journal} {{Quantum}}\ }\textbf {\bibinfo {volume}
  {2}},\ \bibinfo {pages} {79} (\bibinfo {year} {2018})}\BibitemShut {NoStop}%
\bibitem [{\citenamefont {McClean}\ \emph {et~al.}(2017)\citenamefont
  {McClean}, \citenamefont {Kimchi-Schwartz}, \citenamefont {Carter},\ and\
  \citenamefont {de~Jong}}]{PhysRevA.95.042308}%
  \BibitemOpen
  \bibfield  {author} {\bibinfo {author} {\bibfnamefont {J.~R.}\ \bibnamefont
  {McClean}}, \bibinfo {author} {\bibfnamefont {M.~E.}\ \bibnamefont
  {Kimchi-Schwartz}}, \bibinfo {author} {\bibfnamefont {J.}~\bibnamefont
  {Carter}}, \ and\ \bibinfo {author} {\bibfnamefont {W.~A.}\ \bibnamefont
  {de~Jong}},\ }\href {\doibase 10.1103/PhysRevA.95.042308} {\bibfield
  {journal} {\bibinfo  {journal} {Phys. Rev. A}\ }\textbf {\bibinfo {volume}
  {95}},\ \bibinfo {pages} {042308} (\bibinfo {year} {2017})}\BibitemShut
  {NoStop}%
\bibitem [{\citenamefont {Li}\ and\ \citenamefont
  {Benjamin}(2017)}]{PhysRevX.7.021050}%
  \BibitemOpen
  \bibfield  {author} {\bibinfo {author} {\bibfnamefont {Y.}~\bibnamefont
  {Li}}\ and\ \bibinfo {author} {\bibfnamefont {S.~C.}\ \bibnamefont
  {Benjamin}},\ }\href {\doibase 10.1103/PhysRevX.7.021050} {\bibfield
  {journal} {\bibinfo  {journal} {Phys. Rev. X}\ }\textbf {\bibinfo {volume}
  {7}},\ \bibinfo {pages} {021050} (\bibinfo {year} {2017})}\BibitemShut
  {NoStop}%
\bibitem [{\citenamefont {Temme}\ \emph {et~al.}(2017)\citenamefont {Temme},
  \citenamefont {Bravyi},\ and\ \citenamefont
  {Gambetta}}]{PhysRevLett.119.180509}%
  \BibitemOpen
  \bibfield  {author} {\bibinfo {author} {\bibfnamefont {K.}~\bibnamefont
  {Temme}}, \bibinfo {author} {\bibfnamefont {S.}~\bibnamefont {Bravyi}}, \
  and\ \bibinfo {author} {\bibfnamefont {J.~M.}\ \bibnamefont {Gambetta}},\
  }\href {\doibase 10.1103/PhysRevLett.119.180509} {\bibfield  {journal}
  {\bibinfo  {journal} {Phys. Rev. Lett.}\ }\textbf {\bibinfo {volume} {119}},\
  \bibinfo {pages} {180509} (\bibinfo {year} {2017})}\BibitemShut {NoStop}%
\bibitem [{\citenamefont {Schwenk}\ \emph {et~al.}(2018)\citenamefont
  {Schwenk}, \citenamefont {Reiner}, \citenamefont {Zanker}, \citenamefont
  {Tian}, \citenamefont {Lepp\"akangas},\ and\ \citenamefont
  {Marthaler}}]{PhysRevA.97.042310}%
  \BibitemOpen
  \bibfield  {author} {\bibinfo {author} {\bibfnamefont {I.}~\bibnamefont
  {Schwenk}}, \bibinfo {author} {\bibfnamefont {J.-M.}\ \bibnamefont {Reiner}},
  \bibinfo {author} {\bibfnamefont {S.}~\bibnamefont {Zanker}}, \bibinfo
  {author} {\bibfnamefont {L.}~\bibnamefont {Tian}}, \bibinfo {author}
  {\bibfnamefont {J.}~\bibnamefont {Lepp\"akangas}}, \ and\ \bibinfo {author}
  {\bibfnamefont {M.}~\bibnamefont {Marthaler}},\ }\href {\doibase
  10.1103/PhysRevA.97.042310} {\bibfield  {journal} {\bibinfo  {journal} {Phys.
  Rev. A}\ }\textbf {\bibinfo {volume} {97}},\ \bibinfo {pages} {042310}
  (\bibinfo {year} {2018})}\BibitemShut {NoStop}%
\bibitem [{\citenamefont {{Kandala}}\ \emph {et~al.}(2018)\citenamefont
  {{Kandala}}, \citenamefont {{Temme}}, \citenamefont {{Corcoles}},
  \citenamefont {{Mezzacapo}}, \citenamefont {{Chow}},\ and\ \citenamefont
  {{Gambetta}}}]{2018arXiv180504492K}%
  \BibitemOpen
  \bibfield  {author} {\bibinfo {author} {\bibfnamefont {A.}~\bibnamefont
  {{Kandala}}}, \bibinfo {author} {\bibfnamefont {K.}~\bibnamefont {{Temme}}},
  \bibinfo {author} {\bibfnamefont {A.~D.}\ \bibnamefont {{Corcoles}}},
  \bibinfo {author} {\bibfnamefont {A.}~\bibnamefont {{Mezzacapo}}}, \bibinfo
  {author} {\bibfnamefont {J.~M.}\ \bibnamefont {{Chow}}}, \ and\ \bibinfo
  {author} {\bibfnamefont {J.~M.}\ \bibnamefont {{Gambetta}}},\ }\href@noop {}
  {\enquote {\bibinfo {title} {{Extending the computational reach of a noisy
  superconducting quantum processor}},}\ } (\bibinfo {year} {2018}),\ \Eprint
  {http://arxiv.org/abs/1805.04492} {arXiv:1805.04492 [quant-ph]} \BibitemShut
  {NoStop}%
\bibitem [{\citenamefont {Peruzzo}\ \emph {et~al.}(2014)\citenamefont
  {Peruzzo}, \citenamefont {McClean}, \citenamefont {Shadbolt}, \citenamefont
  {Yung}, \citenamefont {Zhou}, \citenamefont {Love}, \citenamefont
  {Aspuru-Guzik},\ and\ \citenamefont {O’Brien}}]{peruzzo2014variational}%
  \BibitemOpen
  \bibfield  {author} {\bibinfo {author} {\bibfnamefont {A.}~\bibnamefont
  {Peruzzo}}, \bibinfo {author} {\bibfnamefont {J.}~\bibnamefont {McClean}},
  \bibinfo {author} {\bibfnamefont {P.}~\bibnamefont {Shadbolt}}, \bibinfo
  {author} {\bibfnamefont {M.-H.}\ \bibnamefont {Yung}}, \bibinfo {author}
  {\bibfnamefont {X.-Q.}\ \bibnamefont {Zhou}}, \bibinfo {author}
  {\bibfnamefont {P.~J.}\ \bibnamefont {Love}}, \bibinfo {author}
  {\bibfnamefont {A.}~\bibnamefont {Aspuru-Guzik}}, \ and\ \bibinfo {author}
  {\bibfnamefont {J.~L.}\ \bibnamefont {O’Brien}},\ }\href {\doibase
  10.1038/ncomms5213} {\bibfield  {journal} {\bibinfo  {journal} {Nat.
  Commun.}\ }\textbf {\bibinfo {volume} {5}},\ \bibinfo {pages} {4213}
  (\bibinfo {year} {2014})}\BibitemShut {NoStop}%
\bibitem [{\citenamefont {Kardar}(2007)}]{kardar2007statistical}%
  \BibitemOpen
  \bibfield  {author} {\bibinfo {author} {\bibfnamefont {M.}~\bibnamefont
  {Kardar}},\ }\href
  {http://www.cambridge.org/us/catalogue/catalogue.asp?isbn=9780521873420}
  {\emph {\bibinfo {title} {Statistical Physics of Particles}}}\ (\bibinfo
  {publisher} {Cambridge University Press},\ \bibinfo {year}
  {2007})\BibitemShut {NoStop}%
\bibitem [{\citenamefont {Hinton}(2012)}]{Hinton2012}%
  \BibitemOpen
  \bibfield  {author} {\bibinfo {author} {\bibfnamefont {G.~E.}\ \bibnamefont
  {Hinton}},\ }\enquote {\bibinfo {title} {{A Practical Guide to Training
  Restricted Boltzmann Machines}},}\ in\ \href {\doibase
  10.1007/978-3-642-35289-8_32} {\emph {\bibinfo {booktitle} {Neural Networks:
  Tricks of the Trade}}},\ \bibinfo {editor} {edited by\ \bibinfo {editor}
  {\bibfnamefont {G.}~\bibnamefont {Montavon}}, \bibinfo {editor}
  {\bibfnamefont {G.~B.}\ \bibnamefont {Orr}}, \ and\ \bibinfo {editor}
  {\bibfnamefont {K.-R.}\ \bibnamefont {M{\"u}ller}}}\ (\bibinfo  {publisher}
  {Springer Berlin Heidelberg},\ \bibinfo {address} {Berlin, Heidelberg},\
  \bibinfo {year} {2012})\ pp.\ \bibinfo {pages} {599--619},\ \bibinfo
  {edition} {2nd}\ ed.\BibitemShut {Stop}%
\bibitem [{\citenamefont {{Kingma}}\ and\ \citenamefont
  {{Ba}}(2014)}]{2014arXiv1412.6980K}%
  \BibitemOpen
  \bibfield  {author} {\bibinfo {author} {\bibfnamefont {D.~P.}\ \bibnamefont
  {{Kingma}}}\ and\ \bibinfo {author} {\bibfnamefont {J.}~\bibnamefont
  {{Ba}}},\ }\href@noop {} {\enquote {\bibinfo {title} {{Adam: A Method for
  Stochastic Optimization}},}\ } (\bibinfo {year} {2014}),\ \Eprint
  {http://arxiv.org/abs/1412.6980} {arXiv:1412.6980 [cs.LG]} \BibitemShut
  {NoStop}%
\bibitem [{\citenamefont {Tieleman}(2008)}]{Tieleman:2008:TRB:1390156.1390290}%
  \BibitemOpen
  \bibfield  {author} {\bibinfo {author} {\bibfnamefont {T.}~\bibnamefont
  {Tieleman}},\ }in\ \href {\doibase 10.1145/1390156.1390290} {\emph {\bibinfo
  {booktitle} {Proceedings of the 25th International Conference on Machine
  Learning}}},\ \bibinfo {series and number} {ICML '08}\ (\bibinfo  {publisher}
  {ACM},\ \bibinfo {address} {New York, NY, USA},\ \bibinfo {year} {2008})\
  pp.\ \bibinfo {pages} {1064--1071}\BibitemShut {NoStop}%
\end{thebibliography}%

\appendix

\section{Local Thermalization and QBM Quench Dynamics}\label{sec:qbm_quench_therm}

Let us now show Eq.~\eqref{eq:time_average} in more detail, beginning only from the ETH ansatz of Eq.~\eqref{eq:eth_ansatz}; our construction is based on one presented in~\cite{DAlessio2016}. Once again, consider a quench of the form discussed in Sec.~\ref{sec:quench}, with:
\begin{equation}
    \ket{\psi\left(t\right)}=\sum\limits_i c_i\ce^{-\ci E_i t}\ket{E_i}.
\end{equation}
Then, given an operator $O$, we have that:
\begin{equation}
    \begin{aligned}
        \overline{O}&\equiv\lim\limits_{t\to\infty}\frac{1}{t}\int\limits_0^t\dd{t'}\bra{\psi\left(t'\right)}O\ket{\psi\left(t'\right)}\\
        &=\lim\limits_{t\to\infty}\frac{1}{t}\int\limits_0^t\dd{t'}\sum\limits_{i,j}c_i^* c_j\ce^{-\ci\left(E_j-E_i\right)t'}\bra{E_i}O\ket{E_j}\\
        &=\sum\limits_i\left\lvert c_i\right\rvert^2\bra{E_i}O\ket{E_i}.
        \label{eq:o_time_avg}
    \end{aligned}
\end{equation}
Taking $O$ to have a volume of support $k=o\left(n\right)$ and using the ETH ansatz of Eq.~\eqref{eq:eth_ansatz}, we then have (assuming $S=\Omega\left(n\right)$) that:
\begin{equation}
    \overline{O}=\sum\limits_i\left\lvert c_i\right\rvert^2 O_\omega\left(E_i\right)+\mathcal{O}\left(\ce^{-\frac{n}{2}}\right).
\end{equation}
Defining:
\begin{equation}
E\equiv\bra{\psi\left(0\right)}H\ket{\psi\left(0\right)}=\sum\limits_i\left\lvert c_i\right\rvert^2 E_i,
\end{equation}
we Taylor expand $O_\omega\left(E_i\right)$ about $E$ to find that:
\begin{equation}
    \begin{aligned}
        \overline{O}&=\sum\limits_i\left\lvert c_i\right\rvert^2\left(O_\omega\left(E\right)+\left(E_i-E\right)\left.\dv{O_\omega\left(E'\right)}{E'}\right|_E+\frac{1}{2}\left(E_i-E\right)^2\left.\dv[2]{O_\omega\left(E'\right)}{\left(E'\right)}\right|_E\right)\\
        &+\mathcal{O}\left(\ce^{-\frac{n}{2}}+\frac{\mathbb{E}_i\left[\left\lvert E_i-E\right\rvert^3\right]}{E^3}\right)\\
        &=O_\omega\left(E\right)+\mathcal{O}\left(\ce^{-\frac{n}{2}}+\frac{\mathbb{E}_i\left[\left\lvert E_i-E\right\rvert^2\right]}{E^2}\right).
    \end{aligned}
\end{equation}
Thus, the degree of approximation is good so long as $\frac{\mathbb{E}_i\left[\left\lvert E_i-E\right\rvert^2\right]}{E^2}$ is small.

It is also true that the average difference between $\bra{\psi\left(t\right)}O\ket{\psi\left(t\right)}$ and its long-time average $\overline{O}$ is small~\cite{DAlessio2016}. We calculate using the ETH ansatz of Eq.~\eqref{eq:eth_ansatz} that:
\begin{equation}
    \begin{aligned}
        \lim\limits_{t\to\infty}\frac{1}{t}\int\limits_0^t\dd{t'}\bra{\psi\left(t'\right)}O\ket{\psi\left(t'\right)}^2&=\lim\limits_{t\to\infty}\frac{1}{t}\int\limits_0^t\dd{t'}\sum\limits_{i,j,k,l}c_i^* c_j c_k^* c_l \ce^{-\ci\left(E_j+E_l-E_i-E_k\right)t'}\\
        &\times\bra{E_i}O\ket{E_j}\bra{E_k}O\ket{E_l}\\
        &=\left(\sum\limits_i\left\lvert c_i\right\rvert^2\bra{E_i}O\ket{E_i}\right)^2+\sum\limits_{i\neq j}\left\lvert c_i\right\rvert^2\left\lvert c_j\right\rvert^2\left\lvert\bra{E_i}O\ket{E_j}\right\rvert^2.
    \end{aligned}
\end{equation}
We therefore have from Eq.~\eqref{eq:o_time_avg} and the ETH ansatz of Eq.~\eqref{eq:eth_ansatz} that:
\begin{equation}
    \begin{aligned}
        \lim\limits_{t\to\infty}\frac{1}{t}\int\limits_0^t\dd{t'}\left(\bra{\psi\left(t'\right)}O\ket{\psi\left(t'\right)}^2-\overline{O}^2\right)&=\sum\limits_{i\neq j}\left\lvert c_i\right\rvert^2\left\lvert c_j\right\rvert^2\left\lvert\bra{E_i}O\ket{E_j}\right\rvert^2\\
        &\leq\max_{i\neq j}\left\lvert\bra{E_i}O\ket{E_j}\right\rvert^2\\
        &=\mathcal{O}\left(\ce^{-n}\right),
    \end{aligned}
\end{equation}
where once again we have assumed that $S=\Omega\left(n\right)$.

Thus, assuming that expectation values of $O$ in the microcanonical and canonical ensembles are equivalent up to $\mathcal{O}\left(\frac{k}{n_v}\right)$ terms (which is true for nonintegrable systems when the microcanonical entropy is concave in the energy and the energy is extensive in the system volume)~\cite{TOUCHETTE2004138,Ellis2000,kardar2007statistical,PhysRevE.97.012140,Garrison2018}, we have that:
\begin{equation}
    \bra{\psi\left(t\right)}O\ket{\psi\left(t\right)}=\frac{\tr\left(O\ce^{-\beta\left(\psi\right)H}\right)}{\tr\left(\ce^{-\beta\left(\psi\right) H}\right)}+\mathcal{O}\left(\frac{k}{n_v}+\frac{\mathbb{E}_m\left[\left\lvert E_m-E\right\rvert^2\right]}{E^2}\right)
\end{equation}
for $t$ sufficiently large. When described in the language of the trace distance between the partial traces to a subsystem of size $k$ of $\ket{\psi\left(t\right)}\bra{\psi\left(t\right)}$ and those of a canonical ensemble, this is equivalent to the \emph{subsystem Eigenstate Thermalization Hypothesis}~\cite{PhysRevE.97.012140}.

Thus, all that remains to be shown is that $\frac{\mathbb{E}_m\left[\left\lvert E_m-E\right\rvert^2\right]}{E^2}$ is small; for restricted QBM systems with few hidden units, this is indeed true. Considering the quench procedure described in Sec.~\ref{sec:quench} where we take $\ket{\psi\left(0\right)}$ to be diagonal in the $X$ basis for simplicity, this term is given by~\cite{rigol2008thermalization,DAlessio2016}:
\begin{equation}
    \begin{aligned}
        \frac{\mathbb{E}_m\left[\left\lvert E_m-E\right\rvert^2\right]}{E^2}&=\frac{\bra{\psi\left(0\right)}H^2\ket{\psi\left(0\right)}-\bra{\psi\left(0\right)}H\ket{\psi\left(0\right)}^2}{\bra{\psi\left(0\right)}H\ket{\psi\left(0\right)}^2}\\
        &=\frac{\bra{\psi\left(0\right)}\left(\sum\limits_i b_i \sigma_i^z+\sum\limits_{\upsilon,\eta}w_{\upsilon\eta}\sigma_\upsilon^z \sigma_\eta^z\right)^2\ket{\psi\left(0\right)}}{\bra{\psi\left(0\right)}H\ket{\psi\left(0\right)}^2}\\
        &=\frac{\sum\limits_i b_i^2+\sum\limits_{\upsilon,\eta}w_{\upsilon\eta}^2}{\bra{\psi\left(0\right)}H\ket{\psi\left(0\right)}^2}\\
        &=\mathcal{O}\left(\frac{n+n_v n_h}{n^2}\right).
    \end{aligned}
    \label{eq:therm_condition}
\end{equation}
Therefore, as long as the number of weights is subquadratic in the system size (i.e. $n_v n_h=\mathcal{o}\left(n^2\right)$), then:
\begin{equation}
    \frac{\mathbb{E}_m\left[\left\lvert E_m-E\right\rvert^2\right]}{E^2}=\mathcal{o}\left(1\right).
\end{equation}
Due to the apparent strength of QBMs with small numbers of hidden units (see Sec.~\ref{sec:training} and~\cite{Kieferova2017}), this is not an unreasonable assumption. However, this analysis does not hold for semi-restricted or unrestricted models, and indeed for numerically simulated generic semi-restricted transverse Ising models it seems that this convergence does not hold in the thermodynamic limit (see Sec.~\ref{sec:ergodicity}). On actual training data, though, our QBM/\allowbreak thermometer scheme does seem to train well, even for the semi-restricted transverse Ising model. This could be due to Eq.~\eqref{eq:therm_condition} giving, in general, that the necessary condition for thermalization is:
\begin{equation}
    \sum\limits_{i,j}w_{ij}^2=\mathcal{o}\left(n\right);
\end{equation}
thus, the apparent thermalization of even nonrestricted QBMs in Sec.~\ref{sec:training} may be due to visible-visible couplings being small during training on our considered data distributions. Alternatively, it could be due to our training procedure being robust to even constant errors in estimates of the gradient, and only strongly depending on---for instance---the sign of the gradient. We leave further exploration of this behavior to future work.
%
%

\section{QBM/Thermometer Systems}\label{sec:systems}

In our numerical experiments, we consider the Hamiltonians given in Table~\ref{tab:hams}; in all instances, we take the QBM and thermometer models to be the same. Furthermore, the interaction Hamiltonian between the QBM and thermometer is always of the form of Eq.~\eqref{eq:int_ham}. The restricted XX model is universal for quantum computation~\cite{Cubitt9497}; we restrict the $\sigma_\upsilon^x \sigma_\eta^x$ and $\sigma_\upsilon^z \sigma_\eta^z$ terms to have the same weights (i.e. we consider the XX rather than the XY model) such that the positive phases of the gradient of Eq.~\eqref{eq:u_b_ll} do not vanish when training the $\sigma_\upsilon^x \sigma_\eta^x$ terms.

\begin{table}
    \begin{center}
        \begin{ruledtabular}
            \begin{tabular}{cc}
            Model & QBM/Thermometer Hamiltonian\\
            \hline
            Semi-restricted Transverse Ising Model & $H_{\textrm{QBM/therm}}\left(\bm{\theta}\right)=\sum\limits_i \varGamma_i \sigma_i^x+\sum\limits_i b_i \sigma_i^z+\sum\limits_{\upsilon,i}w_{\upsilon i}\sigma_\upsilon^z \sigma_i^z$\\
            Restricted Transverse Ising Model & $H_{\textrm{QBM/therm}}\left(\bm{\theta}\right)=\sum\limits_i \varGamma_i \sigma_i^x+\sum\limits_i b_i \sigma_i^z+\sum\limits_{\upsilon,\eta}w_{\upsilon\eta}\sigma_\upsilon^z \sigma_\eta^z$\\
            Restricted XX Model & $H_{\textrm{QBM/therm}}\left(\bm{\theta}\right)=\sum\limits_i \varGamma_i \sigma_i^x+\sum\limits_i b_i \sigma_i^z+\sum\limits_{\upsilon,\eta}w_{\upsilon\eta}\left(\sigma_\upsilon^x \sigma_\eta^x+\sigma_\upsilon^z \sigma_\eta^z\right)$
            \end{tabular}
        \end{ruledtabular}
        \caption{The various models considered in our numerical experiments.}
        \label{tab:hams}
    \end{center}
\end{table}

We chose the evolution times for our QBM/\allowbreak thermometer combination uniformly from $\left[\sqrt{\frac{2}{\cpi}},10\sqrt{\frac{2}{\cpi}}\right]$; this defines the energy scale for the model parameters. In these units, we initialized the interaction between QBM and thermometer to be drawn from $N\left(0,1\right)$ (that is, the normal distribution with mean $0$ and variance $1$). For all considered models, we drew $\varGamma$ from $N\left(\overline{\varGamma},2.5\times 10^{-5}\right)$, where $\overline{\varGamma}=1$ for the results in Sec.~\ref{sec:training}. Furthermore, for both the RBMs and the QBMs, we initialized the visible biases $b_\upsilon$ to:
\begin{equation}
    b_\upsilon^{\textrm{init}}=\ln\left(\frac{p_{b_\upsilon}^{\textrm{init}}}{1-p_{b_\upsilon}^{\textrm{init}}}\right),
\end{equation}
where
\begin{equation}
    p_{b_\upsilon}^{\textrm{init}}=\frac{\mathbb{E}_{\bm{d}\sim p_{\textrm{data}}}\left[d_i\right]+1}{2},
\end{equation}
and the initial hidden biases $b_\eta$ were sampled from $N\left(0,2.5\times 10^{-5}\right)$. Finally, the initial weights $w_{ij}$ were sampled from $N\left(0,10^{-4}\right)$. The chosen initial values for the biases and weights were inspired by~\cite{Hinton2012}.

For the results described in Sec.~\ref{sec:ergodicity}, we considered the same QBM/\allowbreak thermometer interaction strength and thermometer parameters as in our training results, but estimated the final trained biases and weights for the QBM to be drawn from $N\left(0,1\right)$.

\section{Training Procedure}\label{sec:training_procedure}

We trained each model using the Adam algorithm~\cite{2014arXiv1412.6980K}, with the hyperparameters $\beta_1=0.5$, $\beta_2=0.9$, and $\epsilon=10^{-8}$. We summarize the learning rates $\alpha$ used for all models in Table~\ref{tab:lrs}; we used the optimal $\alpha$ for each model found via grid search. We estimated $\pdv{\beta\left(\bm{\theta}\right)}{\theta}$ as appearing in Eq.~\eqref{eq:beta_correction} by estimating the difference in $\beta$ in between training steps and dividing by the estimated $\partial_\theta\mathcal{L}\left(\bm{\theta}\right)$ between training steps. We trained the restricted Boltzmann machine using persistent contrastive divergence~\cite{Tieleman:2008:TRB:1390156.1390290}, with a number of persistent chains equal to the size of a mini-batch with $1$ step of Gibbs sampling for both the visible and the hidden layers~\cite{Hinton2012}. When training our QBM/\allowbreak thermometer combination, we randomly averaged observables over $\left\lvert\mathcal{T}\right\rvert=2$ quench evolution times drawn from $\left[\sqrt{\frac{2}{\cpi}},10\sqrt{\frac{2}{\cpi}}\right]$ in the units described in Sec.~\ref{sec:ergodicity} and Appendix~\ref{sec:systems}. We trained each model with a mini-batch size of $16$ over $40$ epochs of $512$ data points each. The final empirical KL divergence and AIC was estimated over $1024$ samples for each model. For the results described in Sec.~\ref{sec:ns}, we used $1$ hidden unit for all models; we saw no significant improvement in training performance with more hidden units for any model, probably due to the simplicity of the data distributions. For the results in Sec.~\ref{sec:ns}, we considered $T_1=T_\phi=75$ in the units described in Sec.~\ref{sec:ergodicity} and Appendix~\ref{sec:systems}. Finally, we simulated estimating each observable $O$ through $\nu=1000$ samples by adding Gaussian noise with a variance of $\frac{\left\langle O^2\right\rangle-\left\langle O\right\rangle^2}{\nu}$.

\begin{table}
    \begin{center}
        \begin{ruledtabular}
            \begin{tabular}{cc}
            Model & Learning Rate $\alpha$\\
            \hline
            Restricted Boltzmann Machine & $1.25\times 10^{-3}$\\
            Exact QBM, Semi-restricted Transverse Ising Model & $4\times 10^{-3}$\\
            Exact QBM, Restricted Transverse Ising Model & $2.25\times 10^{-3}$\\
            Exact QBM, Restricted XX Model & $3\times 10^{-3}$\\
            QBM/Thermometer Combination, Semi-restricted Transverse Ising Model & $2\times 10^{-3}$\\
            QBM/Thermometer Combination, Restricted Transverse Ising Model & $2.25\times 10^{-3}$\\
            QBM/Thermometer Combination, Restricted XX Model & $5\times 10^{-4}$
            \end{tabular}
        \end{ruledtabular}
        \caption{The various learning rates used in our numerical experiments.}
        \label{tab:lrs}
    \end{center}
\end{table}

%
%

\end{document}